\documentclass[fleqn,usenatbib]{mnras}

\usepackage{newtxtext,newtxmath}
\usepackage{hyperref}

\usepackage[T1]{fontenc}
\usepackage{ae,aecompl}

\usepackage{graphicx}	
\usepackage{amsmath}	
\usepackage[bottom, hang, flushmargin]{footmisc}
\usepackage{threeparttable}
\usepackage{xspace}
\usepackage{adjustbox}
\usepackage[caption=false]{subfig}

\newcommand{\aox}{$\alpha_{\rm{ox}}$\xspace}	
\newcommand{\daox}{$\Delta\alpha_{\rm{ox}}$\xspace}

\newcommand{\Lopt}{$L_{2500}$\xspace}
\newcommand{\civ}{\,\ion{C}{iv}\xspace}
\newcommand{\civew}{\,\mbox{\ion{C}{iv} EW}\xspace}
\newcommand{\heii}{\,\ion{He}{ii}\xspace}
\newcommand{\heiiew}{\,\mbox{\ion{He}{ii} EW}\xspace}
\newcommand{\Chandra}{\,\emph{Chandra}\xspace}

\newcommand{\angstrom}{\text{\normalfont\mbox{\AA}}\xspace}
\newcommand{\xray}{\,\hbox{X-ray}\xspace}
\newcommand{\highL}{\hbox{High-$L$}\xspace}

\newcommand*{\thead}[1]{\multicolumn{1}{c}{ #1}}

\title[\aox--\heiiew--\Lopt relations]{What controls the UV-to-X-ray continuum shape in quasars?}

\author[J. D. Timlin et al.]{
John D. Timlin III,$^{1,2 \thanks{E-mail: jxt811@psu.edu}}$
W. N. Brandt,$^{1,2,3}$
and Ari Laor$^{4}$
\\
$^{1}$Department of Astronomy \& Astrophysics, 525 Davey Lab, The Pennsylvania State University, University Park, PA 16802, USA \\
$^{2}$Institute for Gravitation and the Cosmos, The Pennsylvania State University, University Park, PA 16802, USA \\
$^{3}$Department of Physics, 104 Davey Lab, The Pennsylvania State University, University Park, PA 16802, USA\\
$^{4}$Physics Department, Technion, Haifa 32000, Israel
}

\date{Accepted 2021 April 26. Received 2021 April 21; in original form 2021 March 12}

\pubyear{2021}

\begin{document}
\label{firstpage}
\pagerange{\pageref{firstpage}--\pageref{lastpage}}
\maketitle

\begin{abstract}
We present an investigation of the interdependence of the optical-to-\hbox{X-ray} spectral slope (\aox), the \heii equivalent-width (EW), and the monochromatic luminosity at 2500 \AA\ (\Lopt). The values of \aox and \heiiew are indicators of the strength/shape of the quasar ionizing continuum, from the ultraviolet (UV; 1500--2500 \AA), through the extreme ultraviolet (EUV; 300--50 \AA), to the \xray (2 keV) regime. For this investigation, we measure the \heiiew of 206 radio-quiet quasars devoid of broad absorption lines that have high-quality spectral observations of the UV and \hbox{2 keV} \hbox{X-rays}. The sample spans wide redshift ($\approx$ 0.13--3.5) and luminosity (log$($\Lopt) $\approx$ 29.2--32.5 \hbox{erg s$^{-1}$ Hz$^{-1}$}) ranges. We recover the well-known \aox--\Lopt and \heiiew--\Lopt anti-correlations, and we find a similarly strong correlation between \aox and \heiiew, and thus the overall spectral shape from the UV, through the EUV, to the \xray regime is largely set by luminosity. A significant \aox--\heiiew correlation remains after removing the contribution of \Lopt from each quantity, and thus the emission in the EUV and the \hbox{X-rays} are also directly tied. This set of relations is surprising, since the UV, EUV, and \xray emission are expected to be formed in three physically distinct regions. Our results indicate the presence of a redshift-independent physical mechanism that couples the continuum emission from these three different regions, and thus controls the overall continuum shape from the UV to the \xray regime. 
 
\end{abstract}




\section{Introduction}\label{sec:intro}

Quasars emit radiation across nearly the full range of the electromagnetic spectrum, and it is widely accepted that different physical regions in quasars are responsible for the production of the radiation at different wavelengths (e.g.\ \citealt{Elvis1994,Krawczyk2013}). According to the standard quasar depiction (e.g.\ \citealt{Netzer2013}), the \xray emission originates in the accretion-disk corona surrounding the central super-massive black hole, the ultraviolet (UV) through optical emission is typically considered to be largely produced in the accretion disk,\footnote{Though see, e.g. \citet{Lawrence2012}, \citet{Antonucci2015}, \citet{Lawrence2018}, and Section~2.2 of \citet{Davis2020} for observational challenges and alternative models.} and infrared emission is generated by absorption and re-emission from dust at larger scales surrounding the disk. Combining the emission from these regions in the quasar spectral energy distribution (SED) provides a useful picture of the multi-wavelength behavior of quasars that can be used to assess different theoretical models. The extreme UV (EUV) radiation (above 50 eV) largely responsible for the high ionization lines in quasars cannot generally be observed directly due to intervening absorption systems; see \citet{Scott2004}, \citet{Stevans2014}, and \citet{Lusso2015} for composite spectra of the ionizing SED down to $\sim 500$ \AA. This lack of direct observation leaves a wide gap in a critical region of the quasar SED that limits model assessment. 

Different proxies have been used to measure indirectly the strength of the ionizing SED in quasars. One indirect measurement of the amount of ionizing radiation can be determined from the high-ionization emission lines in the quasar spectrum. In particular, the \ion{C}{iv}$\lambda1549$ (hereafter, \civ) and \hbox{\heii$\lambda1640$} (hereafter, \heii) emission lines have been used in previous work as proxies for the amount of ionizing radiation since the formation of these ions requires photons with energies above 47.9 eV and 54.4 eV, respectively. Of these two emission lines, the \civ line has been more extensively studied as it is a prominent line in the rest-frame UV quasar spectrum and thus the line properties (e.g.\ the rest-frame equivalent width; EW) can generally be measured even for low-luminosity or distant quasars. Conversely, the \heii emission line is a weak line that resides in a complex region of a quasar's rest-frame UV spectrum, being surrounded by the red-tail of the \civ line, the \ion{O}{iii]}$\lambda1663$ line, and potentially \ion{Fe}{ii} emission-line blends \citep{Vestergaard2001}. High-quality spectral observations are therefore necessary to obtain robust measurements of the \heii emission-line properties. 

Although the strengths of both \civ and \heii depend on the strength of the EUV ionizing radiation, the line excitation mechanisms are fundamentally different, which can have significant impact on how well they represent the ionizing SED. The \civ ion is produced by ionizing photons above 47.9 eV, but this sets only the abundance of the \civ ion and not the strength of the \ion{C}{iv}$\lambda1549$ line. This line is excited by collisions of the \civ ion with electrons with an energy above 8 eV, and thus this line is also sensitive to the temperature of the gas which, in turn, depends on the gas heating versus cooling rate. The heating rate is set by the mean kinetic energy that incident photons give to the electrons, following an ionization (of typically H and He). The cooling occurs through collisionally excited lines, which mostly originate from metals, such as C, N, O, and in particular the \ion{C}{iv}$\lambda1549$ line. The gas temperature is also sensitive to the gas metallicity (higher metallicity cools the gas, and weakens the \civ line; e.g.\ see Figure~5 of \citealt{Baskin2014}) thus adding another factor that determines the \civ emission-line strength. Radiative-transfer effects in the nuclear environment (e.g.\ \citealt{Chiang1996,Waters2016}) can also cause the \civ emission line, which is a resonance line with potentially large optical depth, to be anisotropically emitted (e.g.\ \citealt{Ferland1992,Goad2014}) which can also affect the observed line strength. Although the \civ line provides a useful first-order measure of the hardness of the ionizing spectrum above \hbox{$\approx$ 50 eV}, which sets the \civ ion abundance, the measured strength of the line can be influenced by other factors.

The \heii line, on the other hand, is produced by recombination of \ion{He}{iii} to \heii, and its strength is therefore set by the ionization rate of \heii to \ion{He}{iii}. In contrast to the \civ emission line, the \hbox{\heii$\lambda1640$} line provides a ``clean" measure of the number of photons above 54.4 eV which are incident on the quasar broad emission-line region gas. The He II line originates from an excited state, with a negligible population, and thus the line is optically thin and not subject to the resonant scattering which may affect the \civ line. Although \heii is a weaker emission line than \civ, and is potentially blended, it is a significantly more accurate measure of the ionizing continuum, and is particularly useful in high signal-to-noise spectra where the line can be well measured. 

Another commonly used indirect measurement of the strength of the ionizing SED in quasars is the optical/UV-to-\hbox{X-ray} power-law spectral slope (\aox). This spectral slope reflects the relative strength of the rest-frame UV radiation (in this work, the \hbox{2500 \AA\ }rest-frame monochromatic continuum luminosity, \Lopt, is utilized) to the \xray radiation strength at rest-frame \hbox{2 keV}, thus spanning the full energy range of the ionizing continuum. Functionally, $\alpha_{\rm ox} = 0.3838\times{\rm{log}}_{10}(L_{\rm 2keV}/L_{\rm 2500})$, where $L_{\rm 2keV}$ and $L_{\rm 2500\AA}$ are the monochromatic luminosities at rest-frame \hbox{2 keV} and \hbox{2500 \AA }, respectively. Although the shape of the SED between these boundaries may be considerably more complex than a simple power-law (e.g.\ \citealt{Scott2004}), the \aox parameter has been found to correlate strongly with the \civew (e.g.\ \citealt{Gibson2008, Timlin2020}), despite the fact that the 2 keV photons do not materially ionize \civ. 

All three of the above proxies are observed to be anti-correlated with \Lopt. In the case of the two emission-line EWs, this the well-known Baldwin effect \citep{Baldwin1977}, where previous studies have demonstrated that \heiiew displays a steeper relationship with \Lopt than \civew (e.g.\ \citealt{Zheng1993, Laor1995, Green1996, Korista1998, Dietrich2002}). This luminosity dependence may, however, be a secondary effect, since it has been demonstrated that the primary dependence is likely on the Eddington ratio (e.g.\ \citealt{Baskin2004, Shemmer2015}) perhaps due to radiation shielding by an increasingly geometrically thick inner accretion disk (e.g.\ \citealt{Luo2015, Ni2018}). Similarly, \aox has been shown to be anti-correlated with \Lopt (e.g.\ \citealt{Steffen2006, Just2007,Lusso2016,Timlin2020,Pu2020}), indicating that the relative amount of ionizing radiation present decreases as quasars become more luminous. 

While both \aox and \heiiew can be considered proxies of of the ionizing radiation in a quasar (see Section~\ref{sec:disc_previous} for further discussion of previous results), there have been no large-scale, systematic investigations of their relationship with each other. This can largely be attributed to the fact that both high-quality \xray and rest-frame UV spectral data are needed to obtain robust measurements of the two parameters. This relationship, however, is yet another important aspect of understanding better the nature of the ionizing SED present in quasars, particularly since the \heiiew probes the number of $\approx 50$ eV photons better than the \civew, and provides a more accurate measure of the ionizing continuum. Furthermore, it remains unclear if the \aox--\Lopt relation depends upon the \heiiew--\Lopt relation (or vice versa) and if the combination of these relations leads to a weaker secondary correlation between \aox and \heiiew, or whether this is the primary correlation, and the other two correlations with \Lopt are only secondary. Addressing these issues can provide needed insight into which physical mechanism(s) control the UV-to-\hbox{X-ray} SED strength/shape in quasars.

To assess the relationships between \aox, \heiiew, and \Lopt, we assembled a large sample of 206 radio-quiet type~1 quasars that have high-quality observations of both \xray emission as well as the rest-frame UV spectrum. We specifically focus upon radio-quiet and non-BAL quasars since the \xray emission from this majority population is less likely to be affected by other factors. Removing radio-loud quasars mitigates a possible jet-linked (or other) contribution to the observed \xray emission, and removing BAL quasars lowers the chances that a quasar is highly \xray absorbed. We draw from three archival quasar catalogs that span a wide range in both luminosity and redshift to create our full sample. These samples were selected due to the overlapping high-quality spectral coverage and sensitive \xray observations, allowing us to measure robustly the \heii emission-line and the \xray properties without being limited by a high fraction of upper-limit measurements. While previous investigations have often stacked quasar spectra to measure the average \heiiew robustly and investigate its relationship with, e.g. \Lopt \citep{Dietrich2002}, the data gathered in this investigation allow us to study these relationships for individual quasars and thus provide us a better understanding of the scatter of the relationships. Furthermore, this investigation is the first large-scale study of the relationship between \aox and \heiiew.

This paper is organized as follows: Section~\ref{sec:sample_selection} presents the archival samples from which we selected quasars as well as the multi-wavelength data-collection strategy. The method used to measure the \heii emission-line properties from the collected spectra is described in Section~\ref{sec:spec_meas}. Section~\ref{sec:correlations} presents the results of the correlation analyses used in this work, and Section~\ref{sec:discussion} outlines results from previous related works and provides a discussion of regarding the interpretation of the results. Furthermore, we present the data used in this work in Appendix~\ref{append:A}, and publish a catalog of \xray properties for 26 quasars with high optical luminosity observered during {\emph{Chandra}} Cycle 13 in Appendix~\ref{append:B}. Finally, we quantify the relationship between \heiiew and \civew in Appendix~\ref{append:C}. Throughout this work, we adopt a flat $\Lambda$-CDM cosmology with \mbox{$H_{0}$ = 70 km s$^{-1}$ Mpc$^{-1}$}, \mbox{$\Omega_M$ = 0.3}, and \mbox{$\Omega_{\Lambda}$ = 0.7}, and we utilize the \Chandra Interactive Analysis of Observations (CIAO; \citealt{Fruscione2006}) version 4.10\footnote{\url{http://cxc.harvard.edu/ciao/releasenotes/ciao_4.10_release.html}} software and CALDB version 4.8.3.\footnote{\url{http://cxc.harvard.edu/caldb/}} 


\section{Sample Selection}\label{sec:sample_selection}

This Section describes the multi-wavelength data sets and analysis methods employed to construct the sample of quasars used to investigate the relationships between \aox, \heiiew, and \Lopt. The three quasar data sets described below were specifically selected because they all have high-quality spectral observations of the rest-frame ultraviolet (UV), which is critically important for performing robust measurements of the weak \heii emission line that lies in a blended region, and they have overlapping, sensitive \xray observations yielding a sample with a high \xray detection fraction. Moreover, these three data sets span both a wide redshift range and a wide luminosity range ($\approx 3$ orders of magnitude) which provides a large dynamic range for our correlation analyses (see Section~\ref{sec:correlations}).

\subsection{High-Luminosity Quasars}

The high-luminosity (hereafter \highL) quasar sample was drawn from two archival investigations of the \xray properties of the most-luminous quasars from the Sloan Digital Sky Survey (SDSS; \citealt{York2000}). First, we obtained 32 quasars from \citet{Just2007}, who performed their investigation on the known quasars through the third data release of SDSS (DR3; \citealt{Schneider2005}). In that work, they targeted 21 objects with snapshot \xray observations during \Chandra Cycle 7 and collected archival \xray observations for the remaining 11 quasars. The second set of sources that comprised our \highL sample consisted of 65 \Chandra snapshot observations of the most-luminous quasars through SDSS DR7 \citep{Schneider2010} taken during \Chandra Cycle 13. The target quasars in this campaign were designed to expand upon the work of \citet{Just2007} and thus were optically selected in a similar manner; therefore, these Cycle 13 quasars can be combined with the quasars from \citet{Just2007} without material selection biases affecting the joint sample. The combination of these two samples yielded 97 high-luminosity quasars that span a wide redshift range ($1.5\leq z \leq 4.5$). 

\begin{figure}
	\includegraphics[width=\columnwidth]{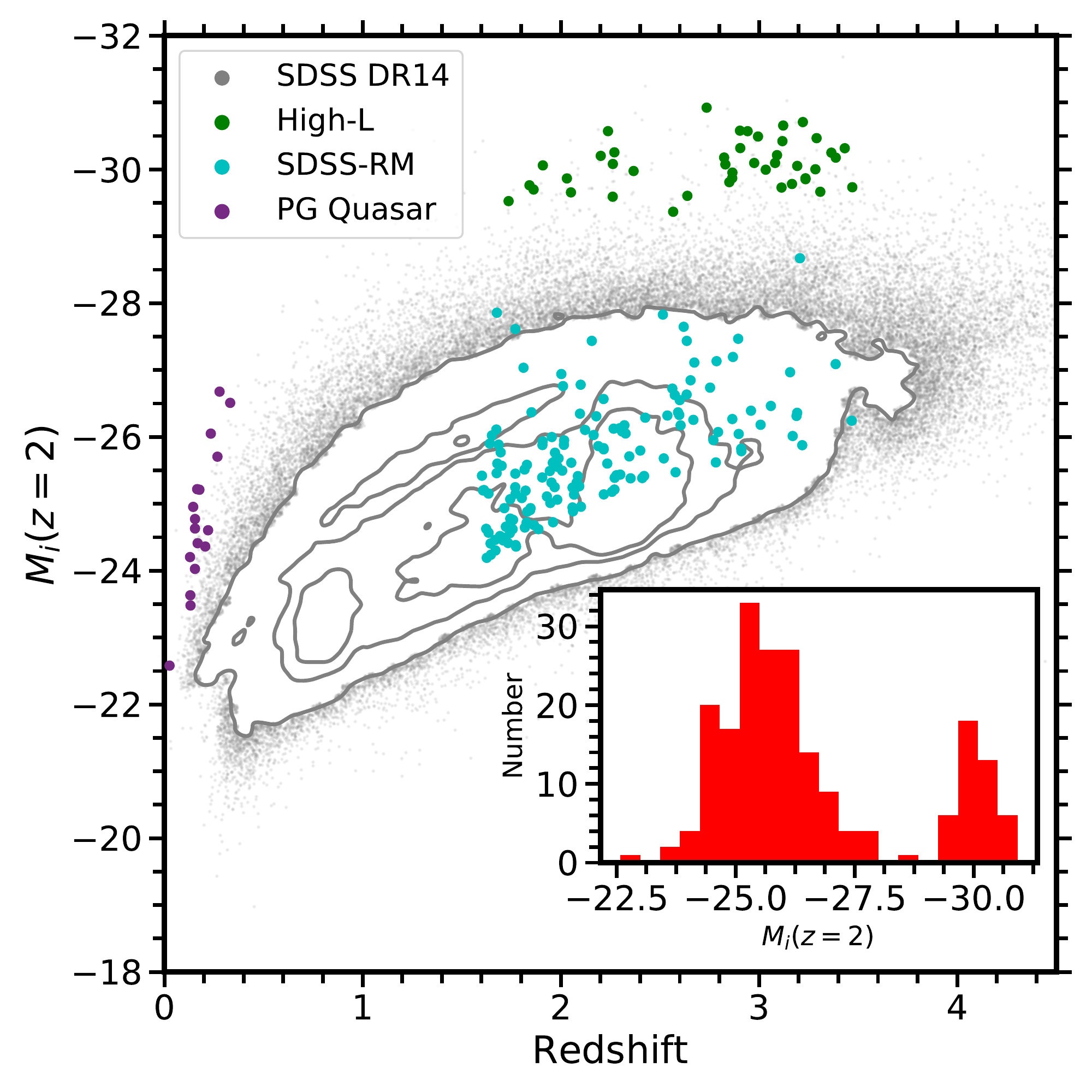}
    \caption{Absolute $i$-band magnitude (corrected to $z=2$; \citealt{Richards2006}) as a function of redshift for the \highL (green), SDSS-RM (blue), and PG (purple) quasars. For reference, the SDSS DR14 quasars \citep{Paris2018} are depicted as grey contours that enclose 35, 68, and 95 percent of the data. The inset plot shows the $M_i(z=2)$ distribution of all of the quasars used to measure \heiiew. The quasars in this sample span a wide range in luminosity at high-$z$, and a wide range in $z$ at low-luminosity, thus allowing us to separate the luminosity and redshift dependences explored here. }
    \label{fig:Mi_z2}
\end{figure}

To generate the \highL sample of quasars for our investigation, these 97 quasars were first required to be radio quiet and to be devoid of BAL features in their rest-frame UV spectra. Furthermore, the redshift range was restricted to $1.6 \leq z \leq 3.5$; the lower limit ensured that a sufficient range of the rest-frame UV continuum was present in the spectrum to determine the presence of a BAL, and the upper limit removed quasars that likely have spectra with low signal-to-noise in which the weak \heii emission line would be difficult to detect. After imposing these restrictions, the final \highL sample consists of 43 quasars (17 from \citealt{Just2007} and 26 from Cycle 13). The absolute magnitude as a function of redshift for these objects is depicted in Figure \ref{fig:Mi_z2} (green points), and some basic properties (e.g.\ number of quasars, mean redshift, and mean absolute magnitude) of the \highL sample are reported in Table~\ref{tab:Samples}. 

This investigation seeks to understand better the relationship between the UV and \xray properties of these quasars. Of particular interest is the \xray-to-optical spectral slope (\aox) which is used as a measure of the hardness of the ionizing SED present in the quasar (e.g.\ \citealt{Timlin2020} and references therein). The \xray data for the quasars in the \highL sample were either directly adopted from \citet{Just2007} (specifically from their Tables~3 and 4) or were measured from the Cycle~13 \Chandra observations. We processed the 26 Cycle~13 \Chandra observations that satisfied our aforementioned cuts and generated an \xray catalog using the same method outlined in Section~3.1 of \citet{Timlin2020} (see Appendix \ref{append:B} for more details). Given the redshift range of the quasars in the \highL sample (median $z = 2.91$), the soft \xray band (\hbox{0.5--2~keV} in the observed frame) measurements, which probe typical rest-frame 1.97--7.81 keV energies, were used to compute the rest-frame monochromatic 2~keV luminosity, $L_{\rm 2keV}$. In cases where a quasar is not detected in this \xray band, we either adopt (in the case of the \citealt{Just2007} objects) or estimate (for the Cycle 13 quasars) the 90\% confidence upper limit \citep{Kraft1991}.\footnote{\xray upper limits in \citet{Just2007} are recorded in their Tables 3 and 4 whereas we compute the upper limit for the Cycle 13 quasars if the observation has a binomial no-source probability of $P_{\rm{B}}>0.01$ (see Section~3.1 and Equation~1 of \citealt{Timlin2020}).} In total, the \highL sample has an \xray detection fraction of $\approx 86\%$ as recorded in Table~\ref{tab:Samples}.

Integral to the calculations and regressions performed in the subsequent sections are reasonable estimates of the uncertainty in the measured parameters. Each parameter is subject to both a measurement uncertainty and uncertainty due to quasar variability. Measurement uncertainties for $L_{\rm 2keV}$ were estimated by propagating the $1\sigma$ errors on the net source counts (computed using the method from \citealt{Lyons1991}) through the conversion from counts to luminosity (see Appendix \ref{append:B} for more details). To this measurement error, we added in quadrature an additional 36\% uncertainty of $L_{\rm 2keV}$ to account for \xray variability, which is adopted from the median-absolute-deviation of the long-timescale \xray variability distribution from \citet{Timlin2020b}. 

The monochromatic luminosity at rest-frame 2500 \AA\ (\Lopt) was computed directly from the well-measured SDSS photometry for each quasar (following the method in \citealt{Richards2006}), and thus the measurement uncertainty does not significantly contribute to the overall uncertainty in \Lopt; rather, the main contribution to the uncertainty in \Lopt comes from variability. For this work, we adopted the long-term root-mean-square variability measurement of 0.2 mag, which corresponds to 20\% of \Lopt, from \citet{MacLeod2010} as the uncertainty due to quasar variability. The total uncertainty in \aox was calculated using standard error propagation with the uncertainties in \Lopt and $L_{\rm 2keV}$.

\begin{table}
\centering
\caption{Summary of quasar samples}
\label{tab:Samples}

\begin{tabular}{l r l l l c}
\hline
Sample & $N_{\rm Q}$ & $f_{\rm detected}^{\rm \heii } $ & $f_{\rm detected}^{\mathrm{X-ray}} $ &  $\langle z \rangle$ &  $ \langle M_i(z=2) \rangle $\\
\thead{(1)} & \thead{(2)} & \thead{(3)} & \thead{(4)} & \thead{(5)} & \thead{(6)} \\
\hline
High-$L$ & 43 & 0.907 & 0.860 & 2.79 & $-$30.07  \\
SDSS-RM & 146 & 0.883 & 1.000 & 2.18 & $-$25.73  \\
PG & 17 & 1.000 & 1.000 & 0.18 & $-$24.77 \\
Total & 206 & 0.898 & 0.971 & \multicolumn{1}{|c|}{--} & \multicolumn{1}{|c|}{--} \\
\hline
\end{tabular}

\begin{flushleft}
\footnotesize{{\it Notes:} Basic properties of the \highL, SDSS-RM, and PG quasar samples. Column (2) reports the number of quasars in each sample, and the \heii emission-line and \xray detection fraction are reported in columns (3) and (4). Columns (5) and (6) present the average redshift and absolute magnitude of each sample. The total number of quasars used to measure correlations is reported in the last row.}
\end{flushleft}

\end{table}

\subsection{SDSS Reverberation-Mapping Quasars}

Another sample of quasars used in this investigation is from the public data archive of the SDSS Reverberation-Mapping Project (SDSS-RM; \citealt{Shen2015}). SDSS-RM was designed to monitor 849 quasars in a $\approx 7$ deg$^2$ field over numerous epochs with the primary goal of measuring time lags between variations in the continuum flux and emission-line flux. SDSS-RM began measuring quasar spectra in 2014 and has been taking data through 2021, accumulating more than 75 spectral epochs of observations. These quasars span a wide range in redshift ($0.1 \leq z \leq 4.5$) and luminosity. Recently, \citet{Shen2019} stacked 32 spectral epochs available through 2014 and publicly released these combined spectra as well as measurements of the continuum and emission-line properties. These high-quality stacked spectra generally have sufficient signal-to-noise to measure robustly even weak lines like the \heii emission line. To include these quasars in our work, the redshift range was restricted as before for the \highL quasars ($1.6 \leq z \leq 3.5$) but their typical luminosity is $\approx 100$ times lower (see Table~\ref{tab:Samples}). Quasars that host BALs in their spectra, as well as radio-loud quasars, were also removed from the sample. Additionally, a magnitude limit of $i \leq 21$ was imposed on these quasars to increase the likelihood that the quasar spectrum had sufficient signal-to-noise to measure the \heii emission-line properties. After these cuts were imposed, 146 SDSS-RM quasars remained in the sample. Figure \ref{fig:Mi_z2} depicts their absolute magnitude as a function of redshift (blue points), and we report the basic sample properties in Table \ref{tab:Samples}. We used the observed flux density at rest-frame 2500 \AA\ from the \citet{Shen2019} catalog to compute \Lopt for these quasars.

The SDSS-RM field has also been covered by \xray observations with {\emph{XMM-Newton}}, and \xray source catalogs are presented in \citet{Liu2020}. The {\emph{XMM-Newton}} observations were performed from 2016--2017 and overlap with 6.13 deg$^2$ of the SDSS-RM field. The source catalogs were generated using only the \xray survey, as opposed to performing forced photometry at the SDSS-RM target positions. We then matched these \xray catalogs to the 849 \hbox{SDSS-RM} quasars. All 146 SDSS-RM quasars that satisfy the sample restrictions above are detected in the soft \xray band in the \citet{Liu2020} catalog (observed frame 0.5--2 keV). Given the median redshift of these 146 quasars ($z = 2.05$), the typical rest-frame band-pass covered by the soft \xray band is 1.52--6.12 keV, and it thus probes rest-frame 2~keV. The reported monochromatic 2~keV flux densities and luminosities, as well as the measurement uncertainties, were adopted for our investigation. As for the \highL sample, we add to the measurement errors an additional 36\% uncertainty in $L_{\rm 2keV}$ to account for \xray variability, and we conservatively adopt the 20\% uncertainty to account for variability of \Lopt.\footnote{We consider this conservative since the SDSS-RM spectra were stacked over multiple epochs, likely averaging out some of the variability contribution.}

\subsection{Palomar-Green Quasars}

The Palomar-Green (PG; \citealt{Schmidt1983}) quasar sample consists of low-to-moderate luminosity, blue type-1 quasars from the PG catalog \citep{Green1986}, many of which reside at lower redshift ($z<0.5$; e.g.\ \citealt{Boroson1992}). Additionally, \citet{Laor2002} gathered 56 PG quasars that had high-quality spectral observations of the rest-frame UV/optical from the {\emph{Hubble Space Telescope}} archive and published their reduced spectra.\footnote{All the reduced spectra are available at \url{http://personal.psu.edu/wnb3/laorbrandtdata/laorbrandtdata.html}} In their sample, 35 PG quasars have coverage of the \heii emission-line region and the surrounding continuum needed for fitting (\hbox{1420--1710 \AA}; see Section~\ref{sec:spec_meas}). Five of these 35 PG quasars were flagged as radio-loud \citep{Kellermann1994} and seven were flagged as either UV absorbed or exhibited a BAL or a miniBAL in the spectrum (e.g.\ \citealt{Brandt2000}). These quasars were removed from our sample leaving 23 quasars in the PG sample. To obtain the 2500~\AA\ luminosity for the PG quasars, we converted the observed flux at rest-frame 3000 \AA\ from \citet{Neugebauer1987} to rest-frame \Lopt, assuming a spectral slope of $\alpha_{\nu}=-1$ in this spectral region. The measurement uncertainties of \Lopt for the PG quasars are generally insignificant compared to the 20\% uncertainty we adopt to account for quasar UV continuum variability.

The PG quasars occupy a significantly lower region in redshift than the quasars in the \highL and SDSS-RM samples, but a similar range of luminosities to the SDSS-RM sample. The low redshift (median $z = 0.165$) implies that measurements of the hard \xray emission (2--10 keV in the observed frame) are required to probe similar rest-frame \xray energies (2.33--11.65~keV) to those used for the other samples. Of the 23 non-BAL, radio-quiet PG quasars, 17 have reported \xray coverage in the 4XMM catalog \citep{Webb2020} or in small-sample archival work (\citealt{Kaspi2005, Piconcelli2005,Inoue2007,Bianchi2009}).\footnote{In cases where the quasar was recorded in both 4XMM and the dedicated archival work, we adopted the value in the archival work to avoid uncertainties that might arise when processing a large data set.} In cases where there were multiple hard \xray observations of a PG quasar, we adopted the earliest observation, to be closest in time to the optical measurement of \citet{Neugebauer1987} used to measure the rest-frame 3000 \AA\ luminosity (repeated observations are available for only three PG quasars). \xray measurement uncertainties were adopted from the \xray catalog that we used to obtain the 2--10~keV flux, and we again incorporate an additional 36\% uncertainty on $L_{\rm 2 keV}$ to account for \xray variability. Uncertainties in \aox are again computed using standard propagation of error. The absolute $i$-band magnitude as a function of redshift for the 17 PG quasars used in this work is depicted in Figure~\ref{fig:Mi_z2} (purple points). 

\subsection{The \heii Sample}

The full sample used in this work was generated by combining the 43 \highL, 146 SDSS-RM, and 17 PG quasars, which yields a total of 206 quasars (hereafter, the \heii sample) that span wide ranges in luminosity and redshift. The \highL and SDSS-RM samples allow us to probe luminosity dependence, independent of redshift, while the SDSS-RM and PG samples allow us to probe redshift dependence, independent of luminosity. All samples have corresponding high-quality spectra from which the \heiiew can be measured. The full catalog of the 206 quasars used in this work is available in machine-readable format as documented in Appendix~\ref{append:A}.


\section{Measuring the \heiiew}\label{sec:spec_meas}

The weak UV \heii emission line peaks at a vacuum wavelength of $\lambda \approx 1640$ \AA\ and is surrounded by \ion{C}{iv}1549\AA\ on the blue side and \ion{O}{iii]}1663\AA\ at redder wavelengths. Fitting the line profile of \heii can be complicated if the spectrum has low signal-to-noise due to the fact that this line is both weak and is in close proximity to these other broad emission lines; however, the quasars in our sample were specifically selected due to their high-quality spectra to mitigate this problem. Moreover, a simpler method of systematically measuring the EW of this line was implemented in our work to avoid uncertainties that might arise due to model fitting of the blended emission-line profiles in this spectral region. The method used in our work is similar to that in \citet{Baskin2013} in which the \heiiew was used as a measure of the ionizing SED hardness. 

In our investigation, the local continuum in the \heii emission-line region of each spectrum was determined by fitting a linear model to the median values in the continuum windows 1420--1460~\AA\ and \mbox{1680--1700~\AA}. These windows are considered to be relatively free of broad emission-line features, and the quasars investigated in this work are devoid of BALs; therefore, the median fluxes in these windows are generally a good estimate of the local continuum flux. A $3\sigma$ clipping method was also incorporated when determining the median value in order to remove the effects of any narrow spikes in these regions. The \heiiew was then measured by integrating directly the continuum-normalized flux in the window 1620--1650~\AA.\footnote{The SDSS spectra are over-sampled by a factor of three (e.g.\ \citealt{Smee2013}); therefore, we binned the SDSS spectra using the average of every three pixels to avoid issues with correlated noise.} This wavelength range represents the location in the spectrum where the \heii emission maximally contributes to the spectral flux above the red tail of the \ion{C}{iv} emission line and is sufficiently blue-ward of any significant contribution from \ion{O}{iii]} \citep{Baskin2013}. While this method does not capture the full line EW (recovering $\approx 90$\% compared to model fitting), it measures the emission where \heii dominates the spectrum. A key advantage of this method is that the \heii emission is systematically measured, so any excluded emission from the tails of the line from the EW calculation is systematically excluded, and thus our results robustly recover the differences in the line emission between quasars with little bias.

Measurement uncertainties for the \heiiew were determined using a Monte Carlo re-sampling routine with 1000 iterations. This Monte Carlo approach consisted of adding to the observed spectral flux a random value drawn from a normal distribution with mean zero and standard deviation equal to the spectral uncertainty. The spectra from \citet{Laor2002}, however, did not provide the spectral uncertainty per pixel for the PG quasars; therefore, we adopted the standard deviation of the flux in the continuum regions as the spectral uncertainty per pixel. To remain consistent in our analysis of the quasar spectra, this spectral uncertainty was also used for both the PG and SDSS spectra. The \heiiew was computed, as before, for each re-sampled emission line, and the standard deviation of all 1000 iterations was adopted as the $1\sigma$ measurement uncertainty of the \heiiew. In addition to this uncertainty for each quasar, we also incorporated a 10\% uncertainty due to variability in the line emission. We adopted this additional uncertainty value based on the investigation of \citet{Rivera2020}, who measured variability of the \ion{C}{iv} emission line in the SDSS-RM quasars, and found typical variability of the \ion{C}{iv} EW of $\approx$10\%.\footnote{Such an investigation has yet to be performed with the \heii emission line, and would be likely more difficult than for \civ since it is a much weaker line.}

As mentioned before, the \heii line is a weaker broad emission line and thus, on occasion, it cannot be detected above the continuum, even in the high-quality spectra utilized in this investigation. A threshold for line detection was therefore determined and, in cases where the emission line was not detected, an upper limit on the emission-line flux was generated in order to compute upper limits on the \heiiew. The detection threshold was set using the $p(\chi^2, \nu)$ probability commonly employed in \xray astronomy to determine the quality-of-fit of a model to a set of data. This probability estimates the likelihood of the model fit using the $\chi^2$ statistic and the number of degrees of freedom, $\nu$. In this work, the model is considered to be the continuum fit, and the data that are being fit is the spectral measurement in the \heii window 1620--1650 \AA . We consider the \heii emission line to be detected if $p(\chi^2, \nu) \leq 0.001$ (i.e. if the likelihood that the continuum is a reasonable fit to the data is less than $0.1\%$).\footnote{The spectra of the 206 quasars in this work were visually inspected to confirm the choice of this detection threshold.} The \heii emission-line detection fraction for each individual sample in Section \ref{sec:sample_selection}, as well as for the entire \heii sample, is reported in Table~\ref{tab:Samples}. 

If the \heii emission line is not detected significantly above the continuum level, an upper limit on the \heiiew was estimated using a $\Delta\chi^2$ metric (e.g.\ \citealt{FilizAk2012}). To implement this method, we first calculated a reference chi-square value, $\chi^2_{\rm ref}$, between the continuum model and the measured flux in the \heii emission-line region. Then the measured flux in the 1620--1650 \AA\ window is increased uniformly by a constant value, and an updated chi-square value, $\chi^2_{\rm updated}$, was calculated using the new flux and the continuum model. The difference $\Delta\chi^2 = \chi^2_{\rm updated} - \chi^2_{\rm ref}$ was computed and the above process repeated until $\Delta\chi^2  = 6.8$, which corresponds to a 99\% confidence upper limit (e.g.\ \citealt{Press1992}). The \heiiew was then computed, as before, by integrating under the updated emission line that satisfied the $\Delta\chi^2$ threshold. 



\section{Correlations}\label{sec:correlations}

\begin{figure*}
\centering
	\includegraphics[width=0.9\textwidth]{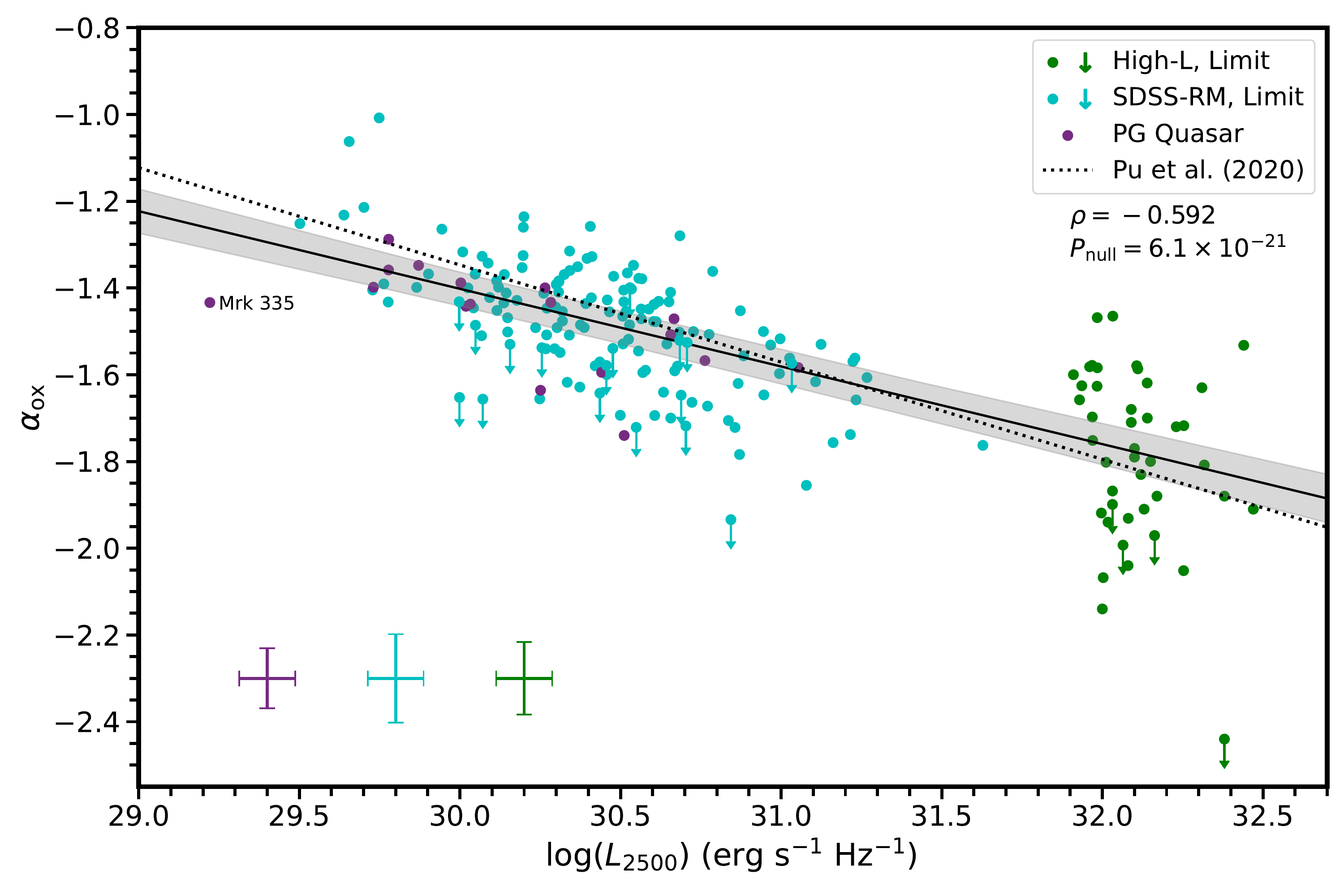}
    \caption{Optical-to-\hbox{X-ray} spectral slope, \aox, as a function of 2500 \AA\ luminosity for the PG (purple), SDSS-RM (blue), and \highL (green) quasars in our sample. In cases where the quasar is not \xray detected (downwards arrows) the 90\% confidence upper limit on \aox is reported. Median error bars are shown in the lower left-hand corner of this panel. The solid black line depicts the best-fit relationship to our data using the \citet{Kelly2007} method and the grey shaded region encloses the $1\sigma$ confidence interval. This fitted relationship is largely consistent with the best-fit relationship from \citet{Pu2020} (black dotted line). The test statistic, $\rho$, and $p$-value from the Spearman rank-order test are reported in the top-right corner (see also Table \ref{tab:Samplestats}). The location of the low-luminosity quasar Mrk 335 is highlighted since this quasar is known to be extremely \xray variable; therefore, the value of \aox is also likely highly variable.  }
    \label{fig:aox_L2500}
\end{figure*}

The primary aim of this work is to investigate the relationships between \aox, \heiiew, and \Lopt. These parameters of interest describe, for each quasar, the nature of the quasar ionizing SED. For \xray unabsorbed quasars, \aox is often used as a direct observational probe of the hardness of the ionizing SED in the quasar. It has been previously shown that there exists an anti-correlation between \aox and \Lopt (e.g., \citealt{Steffen2006, Just2007, Lusso2016, Timlin2020, Pu2020}) indicating that ionizing radiation is less abundant for more luminous quasars. The \heiiew is a more direct indicator of the strength of the ionizing radiation that interacts with the quasar broad emission-line region (see Section~\ref{sec:intro}). Previous investigations have demonstrated that the \heiiew exhibits a strong Baldwin effect having one of the strongest anti-correlations with \Lopt of any broad emission line (e.g.\ \citealt{Zheng1993, Laor1995, Green1996, Korista1998, Dietrich2002}). The following investigation seeks to determine whether a real and significant relation exists between \aox and \heiiew, both of which are proxies of the strength of the ionizing SED in quasars, independent of their individual relationships with \Lopt.

\begin{table*}
\centering
\caption{Statistical tests for correlations and of similarity in each parameter space}
\label{tab:Samplestats}

\begin{tabular}{l r r r r c r c}
\hline
 & & & &\multicolumn{2}{c}{Spearman rank-order} & \multicolumn{2}{l}{Kendall Partial $\tau$} \\ \cline{5-6} \cline{7-8}
\thead{Parameter space} & \thead{Slope ($\sigma_{\rm slope}$)} & \thead{Intercept ($\sigma_{\rm intercept}$)} & \thead{$\sigma_i (\sigma)$} & \thead{$\rho \ (\sigma_{\rho})$} &  \thead{$P_{\rm null}^{\rho}$} & \thead{$\tau \ (\sigma_{\tau})$} &  \thead{$P_{\rm null}^{\tau}$}  \\ 
\thead{(1)} & \thead{(2)} & \thead{(3)} & \thead{(4)} & \thead{(5)} & \thead{(6)} & \thead{(7)}& \thead{(8)}  \\

\hline
\aox--\Lopt & $-$0.179(0.013) & 3.968(0.423) & 0.112(0.048)& $-$0.592(0.029) & $6.1\times10^{-21}$ &\multicolumn{1}{|c|}{--$^{\rm a}$}&\multicolumn{1}{|c|}{--}  \\
\heiiew--\Lopt & $-$0.306(0.021) & 10.038(0.647) & 0.180(0.068) & $-$0.637(0.026) &$4.2\times10^{-24}$ &\multicolumn{1}{|c|}{--}&\multicolumn{1}{|c|}{--} \\
\aox--\heiiew & 0.557(0.039) & $-$1.892(0.027) & 0.080(0.044)& 0.558(0.036) & $2.9\times10^{-18}$&\multicolumn{1}{|c|}{--}&\multicolumn{1}{|c|}{--} \\
\daox--\Lopt & 0.017(0.013) & 0.537(0.406) &0.110(0.046) & $-$0.115(0.043) &$9.9\times10^{-2}$ &\multicolumn{1}{|c|}{--}& \multicolumn{1}{|c|}{--} \\
\daox--\heiiew & 0.187(0.037) & $-$0.115(0.026) & 0.099(0.045) &$-$0.207(0.047) & $2.9\times10^{-3}$ & \multicolumn{1}{|c|}{--} & \multicolumn{1}{|c|}{--} \\
\daox--$\Delta$\heiiew & 0.462(0.060) & $-$0.003(0.009) & 0.074(0.040) &0.285(0.050) & $3.3\times10^{-5}$ & 0.332(0.039) & 1.8$\times10^{-15}$ \\

\hline
\end{tabular}

\begin{flushleft}
\footnotesize{{\it Notes:} Best-fit parameters and correlation statistics for each parameter space in this work. Columns (2)--(4) report the slope, intercept, and intrinsic dispersion along with their $1\sigma$ uncertainties (in parentheses) measured using the method of \citet{Kelly2007}. Furthermore, the correlation coefficient (and $1\sigma$ uncertainty) and $p$-value of the Spearman rank-order test are presented in columns (5)--(6). Uncertainties in the test statistic were computed using a Monte Carlo method (e.g.\ see \citealt{Timlin2020}). Columns (7)--(8) report the test statistic (and $1\sigma$ uncertainty) and $p$-value of a Kendall partial $\tau$ test for the residual parameter spaces in the last two rows. \\
$^{\rm a}$ No partial correlations can be computed for these relationships since there are only two parameters of interest. }
\end{flushleft}

\end{table*}

\begin{figure*}
\centering
	\includegraphics[width=0.9\textwidth]{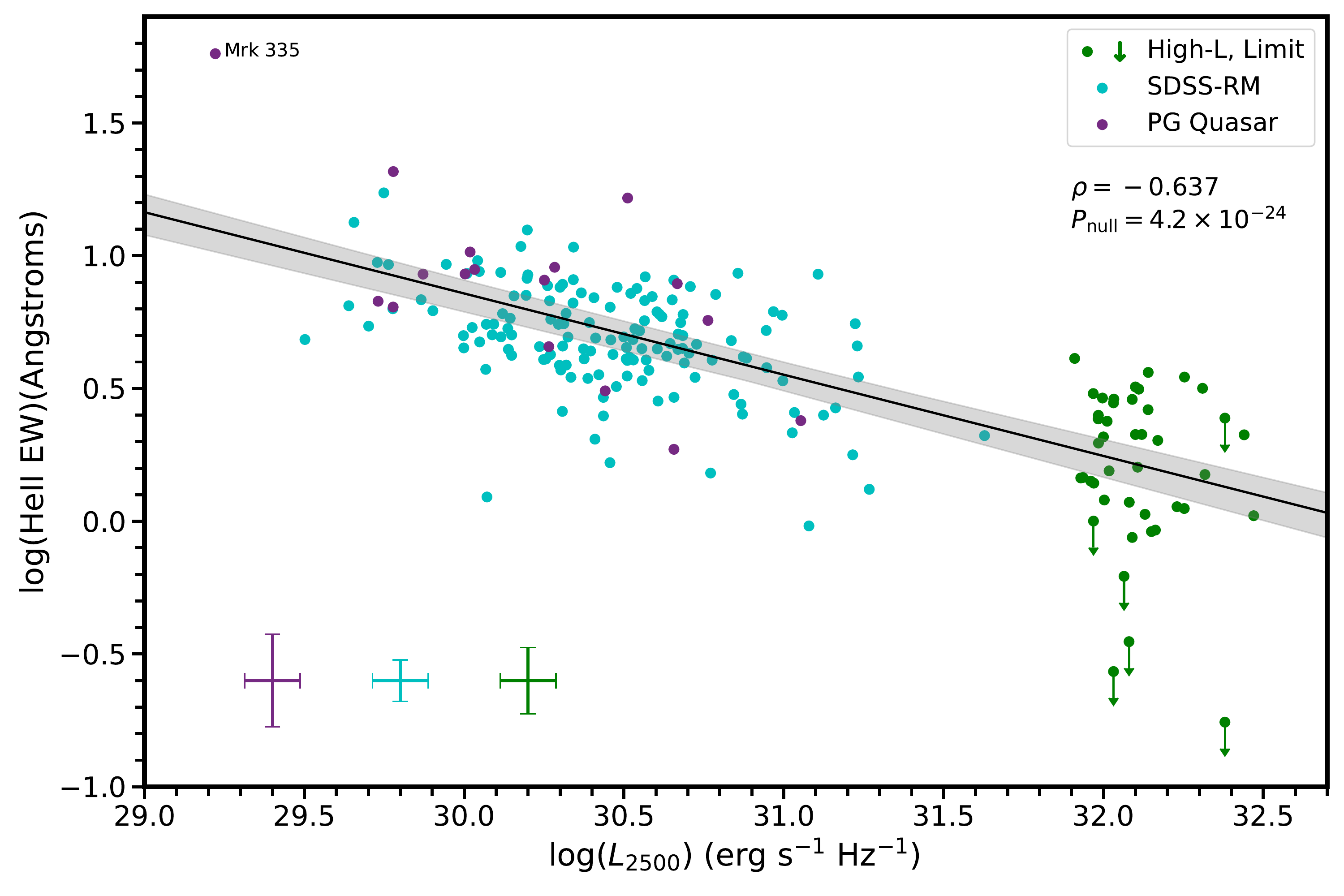}
    \caption{The Baldwin effect between log(\heiiew) and log(\Lopt) for the quasars in our sample. The color scheme is the same as in Figure \ref{fig:aox_L2500}, the black line and grey shaded region depict the best-fit line and the $1\sigma$ confidence interval derived using the method of \citet{Kelly2007}, and the median error bars are shown in the bottom-left corner. The downward-pointing arrows depict the upper limit of the \heiiew when the emission line is not detected with 99\% significance (see Section \ref{sec:spec_meas} for details). As is the case generally for quasar emission lines, a strong Baldwin effect exists in this parameter space according to a Spearman rank-order test (the $\rho$ and $p$-value reported in the upper right-hand corner).}
    \label{fig:heii_L2500}
\end{figure*}

\subsection{The relationships between \aox, \heiiew, and \Lopt}\label{sec:aox_heiiew_corr}

\begin{figure*}
\centering
	\includegraphics[width=0.9\textwidth]{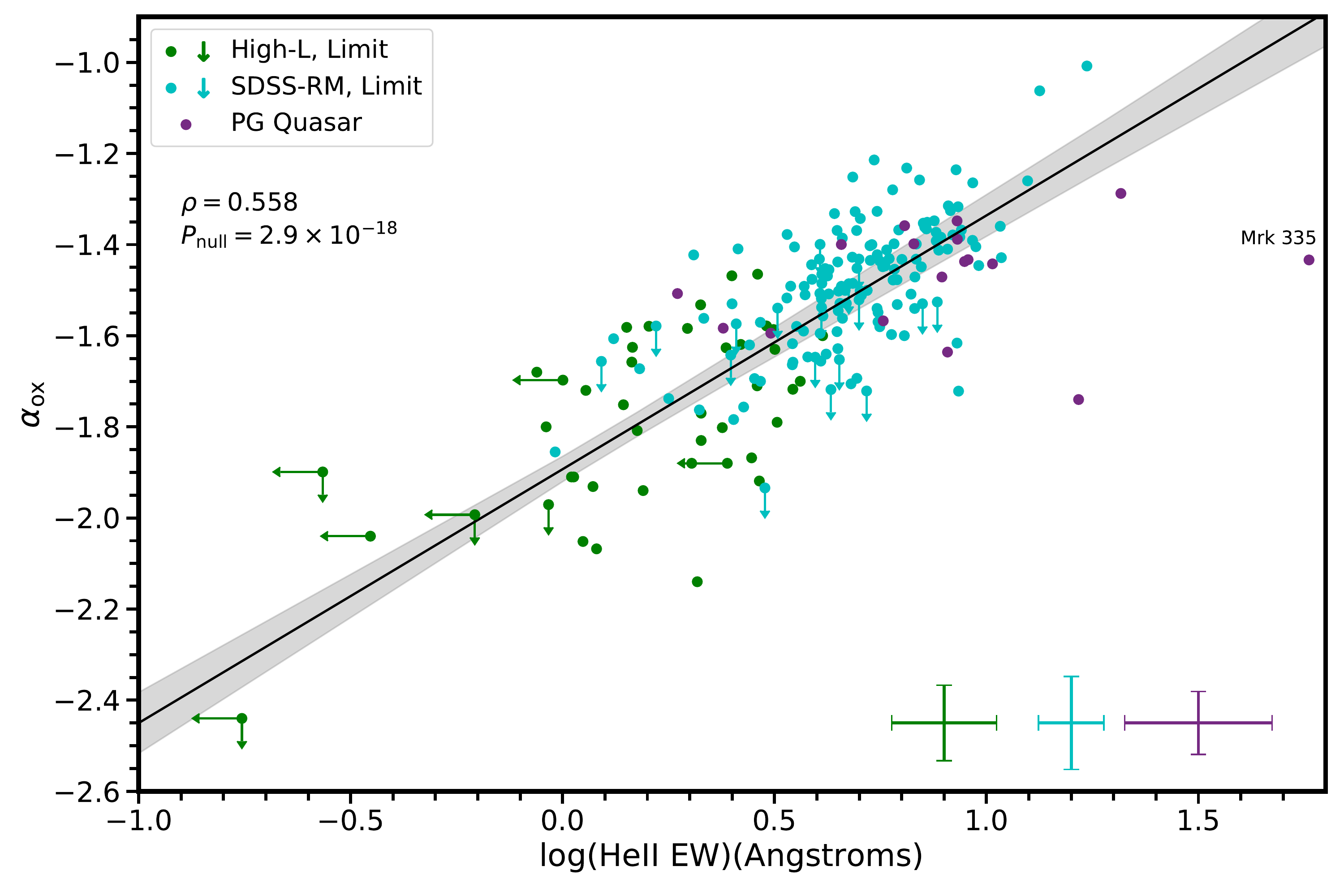}
    \caption{Dependence of \aox on log(\heiiew) for our sample along with the best-fit line and $1\sigma$ confidence interval (color schemes are the same as in Figure \ref{fig:aox_L2500}). Left-pointing arrows depict upper limits in the measurement of the \heiiew, and downward-pointing arrows depict the 90\% confidence upper limit of \aox for \xray non-detections. A significant relationship exists according to a Spearman rank-order test (statistics reported in the top left corner). The three correlations presented in Figures \ref{fig:aox_L2500}--\ref{fig:aox_HeII} are measured to be similarly strong (see Table \ref{tab:Samplestats}).}
    \label{fig:aox_HeII}
\end{figure*}

To assess the joint relationships between the three parameters of interest, we must first understand better the behavior of the data in their two-dimensional parameter spaces (\aox--\Lopt, \heiiew--\Lopt, and \aox--\heiiew).\footnote{In this work, we perform our analyses with respect to $\rm{log}$(\heiiew) and $\rm{log}$(\Lopt); however, for simplicity, we will drop the logarithm notation in the text.} Using the 206 quasars in the \heii sample assembled in Section~\ref{sec:sample_selection}, we first investigated the dependence of \aox on \Lopt, depicted in Figure \ref{fig:aox_L2500}. As mentioned above, there is a well-known anti-correlation between these two parameters with a slope between $\approx 0.14$--$0.22$ depending on the range of \Lopt probed and the properties of the quasars utilized in the investigation (e.g.\ whether the sample included or excluded potentially \xray absorbed quasars; e.g.\ \citealt{Pu2020}). Figure~\ref{fig:aox_L2500} depicts the \aox--\Lopt parameter space for our sample of quasars from the \highL (green points), SDSS-RM (blue points), and PG (purple points) samples. Quasars that were not detected in the \xray are marked by downward-pointing arrows, and are positioned at the location of the 90\% confidence upper limit on \aox.  Median error bars for each sample are depicted in the bottom left of Figure~\ref{fig:aox_L2500}, where the uncertainties are the quadrature sum of both measurement uncertainties and the uncertainties due to variability.\footnote{We highlight the low-luminosity PG quasar Mrk 335 in Figures \ref{fig:aox_L2500}--\ref{fig:aox_HeII} because it has displayed extreme \xray variability on short timescales (e.g.\ \citealt{Komossa2020}), and thus the measurement of \aox, \heiiew, and \Lopt  might be highly variable. This one quasar, however, does not affect greatly the fitted relationships.} 

A Spearman rank-order correlation test was used to determine the significance of the visually apparent anti-correlation in these data; however, this test is not suitable for censored data. To incorporate the censored data points (in this case, the \xray upper limits) into the test, we randomly re-sampled their values from the probability density function (PDF) of the corresponding data set (i.e.\ upper limits in the \highL sample were redrawn from the PDF of the \highL sample only), and the maximum value allowed to be drawn is set at the upper-limit value. In cases where the upper-limit value is smaller than all of the detections in the sample, the re-sampled value is drawn from a uniform distribution between the limit value and the limit value minus the sample's median $1\sigma$ measurement uncertainty. This upper-limit resampling was performed 100 times, and the median test statistic, $\rho$, and corresponding $p$-value of the Spearman test are reported in Table~\ref{tab:Samplestats}.\footnote{We perform the analysis in this manner rather than using the ASURV package (e.g.\ \citealt{Feigelson2014}) to remain consistent with the analysis in the next Sections, where censored data are present in two dimensions and thus cannot be properly analyzed with ASURV. Furthermore, ASURV incorporates the censored data into its algorithms by assuming that the distribution of the censored data is the same as the distribution of the uncensored data. This is not an appropriate assumption for our data, and thus we restricted the upper limit of the censored data point to be the limit value. Thankfully, we have minimized the numbers of upper limits in our samples, and thus our results are not sensitive to details of the upper-limit treatment.} The resulting test statistic ($\rho=-0.592$) and corresponding probability ($p=6.1\times10^{-21}$) that the correlation happens by chance clearly indicate that there is a significant anti-correlation between \aox and \Lopt in this data set. Also reported in Table \ref{tab:Samplestats} is the $1\sigma$ uncertainty value of the test statistic which was estimated using 1000 Monte Carlo re-sampling iterations.

Since a significant anti-correlation exists, a relationship was fitted between these two parameters using the method from \citet{Kelly2007}, as implemented in the {\tt{linmix}} Python package.\footnote{\url{https://linmix.readthedocs.io/en/latest/index.html}} This implementation utilizes a hierarchical Bayesian model to fit a univariate model to data, incorporates errors in both the x- and y-dimensions, and can model censored data in the dependent variable. The best-fit slope ($\beta$) and intercept ($\alpha$) are output as well as an estimate of the intrinsic scatter ($\sigma_{\epsilon}$) around the regression line. This code also reports the $1\sigma$ uncertainty in the fitted parameters and returns the confidence interval of the fit (in this work, we report the $1\sigma$ confidence interval, unless otherwise noted). We depict the best-fit model in Figure~\ref{fig:aox_L2500} as the solid black line, and the confidence interval as the grey shaded region. The slope and intercept of the fitted model are $\beta = -0.179 \pm 0.013$ and $\alpha = 3.968 \pm 0.423$, respectively, and the standard deviation of the intrinsic scatter is $\sigma_{\epsilon} = 0.112 \pm 0.048$ (see Table~\ref{tab:Samplestats}). This fitted relationship is consistent with other values found in the literature (e.g.\ \citealt{Steffen2006, Just2007, Timlin2020}); however, it is slightly flatter than that found most recently in \citet{Pu2020} (dashed line in Figure~\ref{fig:aox_L2500}), which removed \xray absorbed quasars from their analysis. The consistency of this result with those in the literature further confirms that the quasar sample in this work is representative of the general quasar population. 

Next, we investigated the Baldwin effect of the \heiiew for the quasars in the \heii sample. Previous investigations have investigated the \heii Baldwin effect for either small samples or using stacked spectra (see references at the beginning of this Section). Our investigation, on the other hand, is the first large-scale investigation of the \heii Baldwin effect for individual quasars, allowing us also to assess the amount of scatter present in the relationship. Depicted in Figure \ref{fig:heii_L2500} is the \heiiew as a function of \Lopt separated by the three sub-samples in Section \ref{sec:sample_selection}, using the same color scheme as in Figure \ref{fig:aox_L2500}. A Spearman rank-order test was performed as described above to determine the strength of the observed anti-correlation in this parameter space. As has been found in previous investigations, a significant anti-correlation exists between \heiiew and \Lopt (see Table \ref{tab:Samplestats} for the values and uncertainty measurements). The solid-black line and grey-shaded region in Figure \ref{fig:heii_L2500} again depict the best-fit linear model to the data and the $1\sigma$ confidence interval, respectively, fitted using the method from \citet{Kelly2007}. The best-fit slope from our data ($\beta = -0.306 \pm 0.021$) is slightly larger than what has been found in previous investigations (most recently from \citealt{Dietrich2002} who found $\beta=-0.20 \pm 0.04$ using a similarly wide range of luminosity as our investigation). This small difference between the slopes can likely be attributed to the small \heiiew outliers in the \highL sub-sample which steepen the slope of our fitted relationship (see the green arrows in Figure \ref{fig:heii_L2500}). Additionally, \citet{Dietrich2002} ignore their highest luminosity bin when fitting the Baldwin relationship for their data which effectively flattens the fitted slope (see Figure~7 of \citealt{Dietrich2002}). We report the intrinsic scatter of our data in Table~\ref{tab:Samplestats}. Despite this mild difference in slope, the data gathered in this investigation recover the strong \heii Baldwin effect.

The last two-dimensional parameter space evaluated in this work is the \aox--\heiiew space presented in Figure \ref{fig:aox_HeII}, again using the same symbols and color schemes as in Figure \ref{fig:aox_L2500}.\footnote{There are three objects in the \highL sample that have upper limits on the measurements in both dimensions, and thus the arrows point in both directions. To perform the statistical analysis on these three objects, we re-sampled their values as before in both dimensions before including them in the following analyses.} A strong correlation is found between \aox and \heiiew (see Table~\ref{tab:Samplestats} for the Spearman-test results) and thus we again fitted a linear relation to the data in this parameter space. The regression method from \citet{Kelly2007}, however, is not designed to handle censored data as the independent variable; therefore, we accounted for the six \heiiew upper limits by re-sampling the \heiiew values as was described previously for the Spearman correlation test. Upper limit re-sampling was performed 100 times, with each iteration being fitted separately, resulting in a distribution of best-fit parameters.\footnote{In general, using 100 iterations was found to be a sufficient number to produce a stable distribution of best-fit parameters.} The fit that produced the median slope ($\beta = 0.557 \pm 0.039$; see Table~\ref{tab:Samplestats}) was adopted as the best-fit relationship, and is depicted as the black line in Figure~\ref{fig:aox_HeII} (the grey shaded region depicts the corresponding $1\sigma$ confidence interval). 

As depicted in Figures~\ref{fig:aox_L2500}--\ref{fig:aox_HeII}, the SDSS-RM and PG quasars, which have similar luminosity yet largely different redshift (see Table~\ref{tab:Samples}), overlap in all three parameter spaces indicating that the correlations presented here are independent of redshift.

A comparison of the three Spearman $\rho$ values from Table~\ref{tab:Samplestats} for the two-dimensional parameter spaces discussed above suggests that the correlations are all comparable in strength. This similarity of these correlations suggests that there is an interdependence between \aox, \heiiew, and \Lopt that cannot be untangled through investigation of these two-dimensional parameter spaces alone. A reasonable question that we wish to address in the following sub-section is to what degree do two of these relationships impact the third? For example, both \aox and \heiiew are clearly strongly related to \Lopt, but do the \aox--\Lopt and \heiiew--\Lopt correlations drive the \aox--\heiiew relationship, or is there an \aox--\heiiew relationship independent of \Lopt? Moreover, if the relationships with \Lopt drive the \aox--\heiiew relationship, is one of the \Lopt relationships dominant and the other simply a secondary effect? Below, we jointly analyze all three parameters to understand better how they are related.

\subsection{The interdependence of \aox, \Lopt, and \heiiew}\label{sec:prime_second_eff}
\begin{figure}
\centering
	\includegraphics[width=0.9\columnwidth]{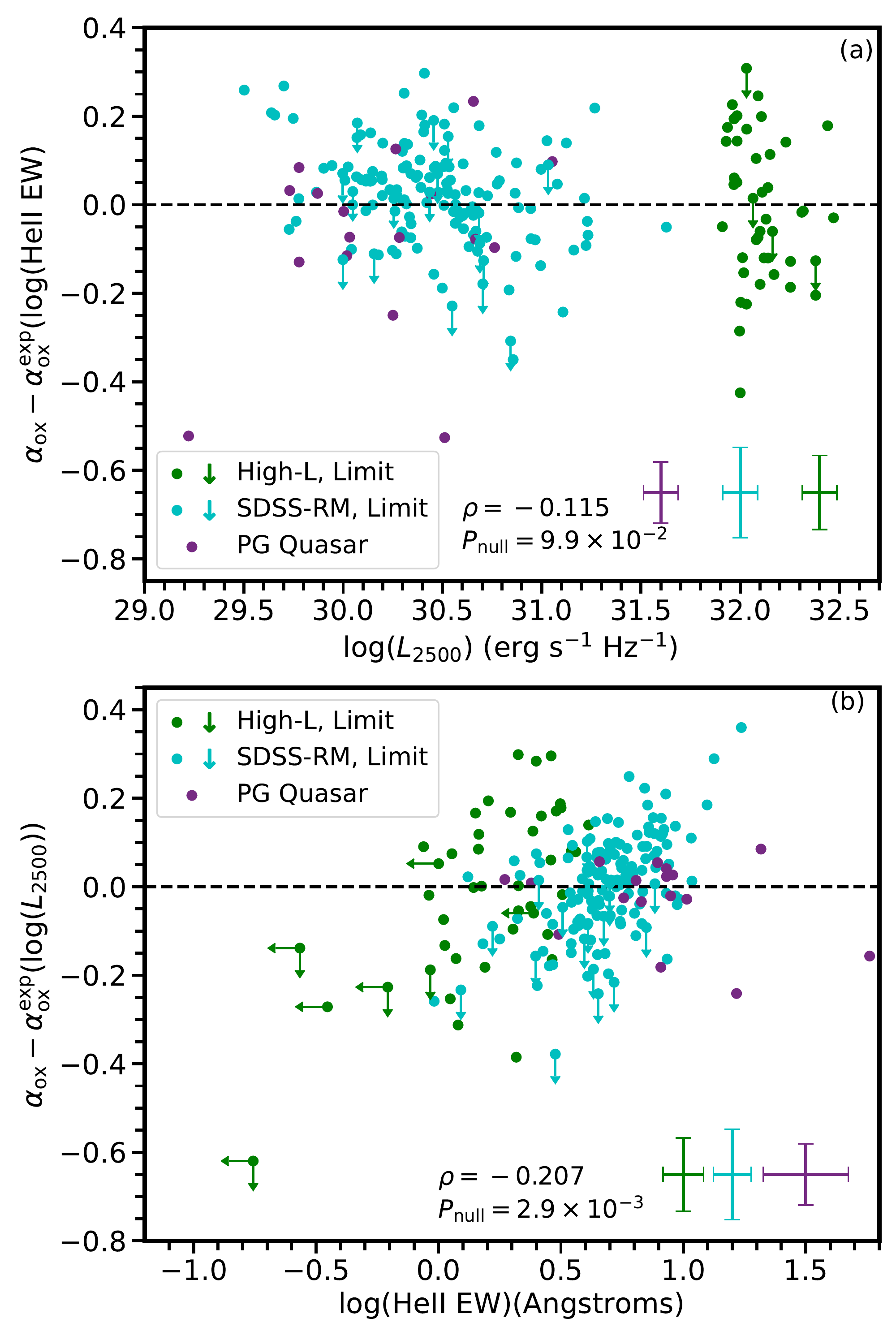}
    \caption{Panel (a): Residual relationship between \aox and log$_{10}$(\Lopt), accounting for the dependence of \aox on log$_{10}$(\heiiew). Panel (b): Residual relationship between \aox and log$_{10}$(\heiiew), accounting for the dependence of \aox on log$_{10}$(\Lopt). The left-pointing arrows in this panel depict the quasars that have upper limits on their measured \heiiew, and the horizontal dashed line represents zero residual. The color scheme and symbols are the same as in Figure \ref{fig:aox_L2500}. In both panels, the Spearman rank-order correlation suggests that, if a correlation exists, it is not highly significant (see Table \ref{tab:Samplestats}). The apparent lack of correlation in these residual parameter spaces, along with the significance of the coefficients in the bivariate regression in Figure \ref{fig:Multivar} suggests that these three parameters might be intrinsically related.}
    \label{fig:Residuals}
\end{figure}

To assess the interdependence of the three parameters of interest, we first investigated residual relationships. In particular, we first examined the residual relationship between the measured \aox and the expected \aox calculated from the relationships shown in Figures \ref{fig:aox_L2500} and \ref{fig:aox_HeII} (we denote the residual as $\Delta\alpha_{\rm ox}$). We then investigated the correlation of $\Delta\alpha_{\rm ox}$ with respect to the parameter not used to compute the residual. For example, in Figure~\ref{fig:Residuals} panel (a), we depict the residual between \aox and the \aox expected from the \aox--\heiiew relationship (Figure~\ref{fig:aox_HeII}) as a function of \Lopt. A Spearman rank-order test performed on the data in this space returns a small correlation coefficient ($\rho = -0.115$, see Table \ref{tab:Samplestats}), which indicates that, if a correlation between $\Delta\alpha_{\rm ox}$ and \Lopt exists, it is weak. Similarly, Figure~\ref{fig:Residuals} panel (b) depicts the residual space with $\Delta\alpha_{\rm ox}$ now being computed between \aox and that expected from the \aox--\Lopt (Figure~\ref{fig:aox_L2500}) relation and depicted as a function of \heiiew. As in the previous case, a Spearman correlation test indicates a potentially mild relationship, if one exists at all ($\rho = -0.207$, see Table~\ref{tab:Samplestats}). The similarity of the correlations for the data in both panels of Figure~\ref{fig:Residuals} suggests that the relationships in Figures~\ref{fig:aox_L2500}--\ref{fig:heii_L2500} are not dominated by a single parameter space. Furthermore, the correlations of the data in Figure~\ref{fig:Residuals} (and Table~\ref{tab:Samplestats}) indicate that \aox correlates as strongly with \heiiew as it does with \Lopt, if not slightly more so, which is a notable result given that previous investigations have been unable to find another parameter that correlates as strongly with \aox as \Lopt (e.g.\ \citealt{Shemmer2008, Liu2021}).

\begin{figure}
\centering
	\includegraphics[width=\columnwidth]{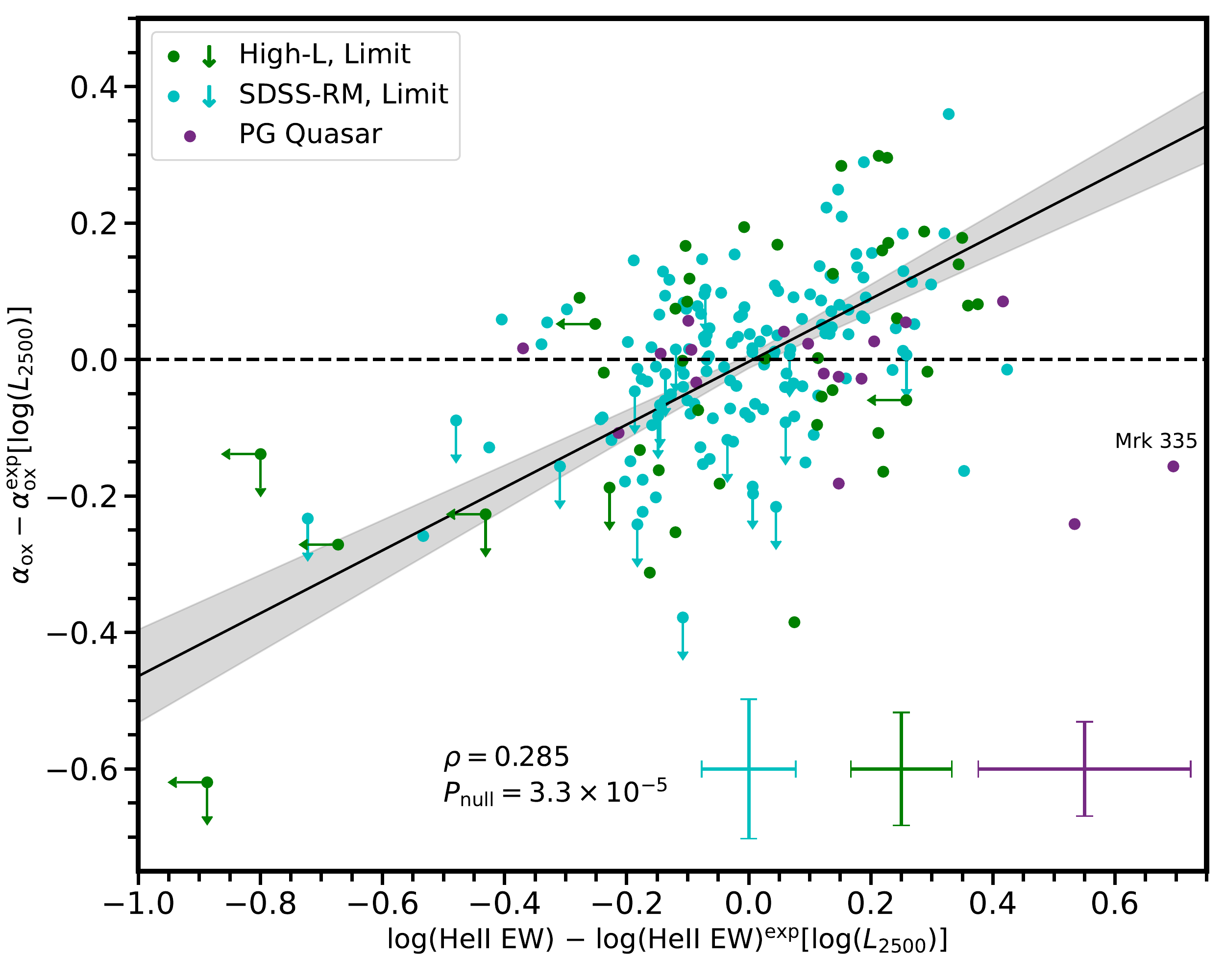}
    \caption{Residual relationship between \aox and \heiiew accounting for both of their dependences on \Lopt. Unlike both panels in Figure~\ref{fig:Residuals}, a significant correlation exists in this parameter space (see Table~\ref{tab:Samplestats}) according to a Spearman test, and thus a relationship was fitted (black solid line) along with the $1\sigma$ confidence interval (grey shaded region). A Kendall partial $\tau$ test further indicates that a correlation exists in this residual space ($\tau = 0.332 \pm 0.039$, $p$-value $= 1.8\times10^{-15}$). The positive correlation between these residuals further demonstrates the existence of a fundamental relationship between \aox and \heiiew after normalizing for their \Lopt dependence. This, in turn, indicates that an additional factor, along with the quasar luminosity, is responsible for setting the ionizing continuum in the quasar. }
    \label{fig:resid-resid}
\end{figure}

\begin{figure*}
\centering
	\includegraphics[width=\textwidth]{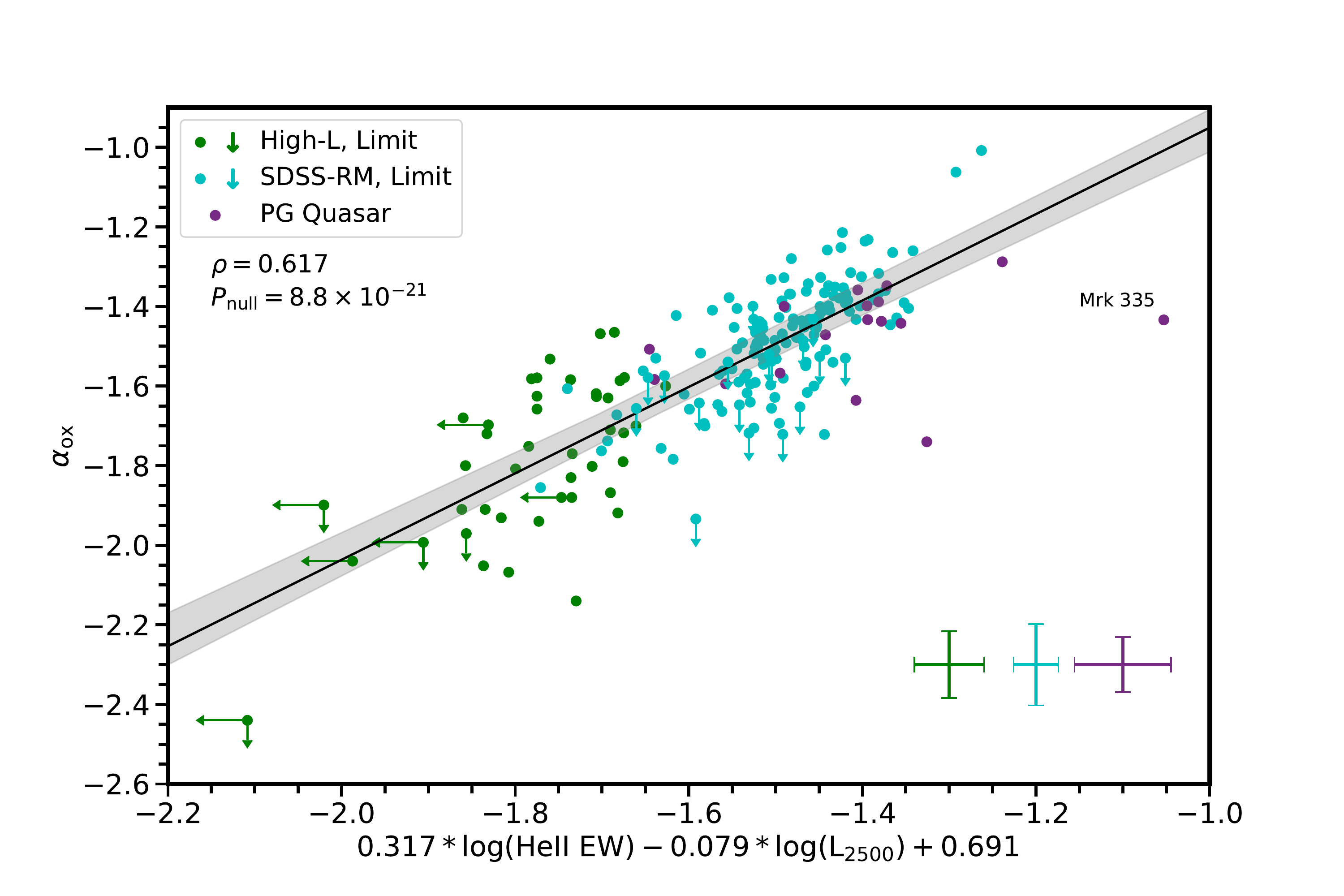}
    \caption{Bivariate regression of \aox as a function of both log(\heiiew) and log(\Lopt) for the quasars in our sample (the colors and symbols are the same as in Figure \ref{fig:aox_L2500}). A strong correlation exists according to the Spearman rank-order test ($\rho$ and the $p$-value are reported in the top left). Furthermore, the coefficients of both log(\heiiew) and log(\Lopt) are statistically different from zero ($7.9\sigma$ and $4.6\sigma$, respectively; see Equation~\ref{eq:multi}), suggesting that both parameters are significantly related to the strength of the ionizing continuum rather than one being a primary and the other only a secondary factor. }
    \label{fig:Multivar}
\end{figure*}

With no single parameter space dominating the relationships in Figures~\ref{fig:aox_L2500}--\ref{fig:aox_HeII}, we next investigated whether a correlation between two parameters exists after removing the contribution of the third parameter. Investigating the correlation between \aox and \heiiew, after accounting for their dependence on \Lopt, is of particular interest in this investigation since these are the probes of the ionizing continuum strength. Figure~\ref{fig:Residuals} panel~(b), in which we already accounted for the luminosity dependence of \aox, shows a stratification of the \highL, RM, and PG quasars which is likely the result of the dependence of \heiiew on \Lopt. To investigate this further, we computed $\Delta$\heiiew by removing the luminosity dependence from the \heiiew using the relationship presented in Figure~\ref{fig:heii_L2500} and the \Lopt value. We depict the $\Delta$\aox--$\Delta$\heiiew residual parameter space in Figure~\ref{fig:resid-resid}, using the same symbols and color scheme as in the previous figures. A Spearman rank-order test indicates that a significant correlation exists in this space; therefore, we fit a relationship to these data using the same method employed to fit the relationship in Figure~\ref{fig:aox_HeII} (see Table~\ref{tab:Samplestats} for the results of the correlation test and the best-fit parameters). 

We confirm the existence of a correlation in the residual space in Figure~\ref{fig:resid-resid} using a Kendall's partial $\tau$ test (e.g.\ see \citealt{Akritas1996} and references therein). Kendall's partial $\tau$ numerically tests for a correlation between two parameters after removing the contribution of their correlations with a third parameter. This test combines the Kendall rank-correlation statistic, $\tau_{12}$ \citep{Kendall1938}, with the Pearson partial product-moment correlation \citep{Kendall1970} resulting in the equation
\begin{equation}
\tau_{12,3} = \frac{\tau_{12} - \tau_{13}\tau_{23}}{[(1-\tau_{13}^{2})(1-\tau_{23}^{2})]^{1/2}},
\end{equation}
where $\tau_{12,3}$ is the partial correlation between parameter 1 and parameter 2, removing the effect of parameter 3. This test is also particularly useful for our investigation since it has been created to handle censored data (which are the upper limits in our case), and it has the capability to provide a significance level to the partial correlation. In our case, we compute $\tau_{\alpha_{\rm ox} {\rm EW}, L_{2500}}$ which is the partial correlation between \aox and \heiiew after removing the effect of \Lopt. We find a partial-correlation strength of $\tau_{\alpha_{\rm ox} {\rm EW}, L_{2500}}=0.332$ with a corresponding $p$-value of $1.8\times10^{-15}$ (see the bottom row of Table~\ref{tab:Samplestats}), which indicates that a significant partial correlation exists in this space. The correlation in this space indicates that there is an intrinsic relationship between \aox and \heiiew that is independent of luminosity. This, in turn, suggests that the \hbox{X-rays} and EUV continuum are coupled through another physical mechanism in addition to the UV luminosity.


The aforementioned tests are robust methods of determining the statistical significance of any residual correlations among the three parameters of interest. Another way to quantify the relationship is through an analysis of the fitting parameters in a bivariate regression. In this work, we fitted \aox as a function of both \heiiew and \Lopt with a bivariate linear model and analyzed the significance, relative to zero, of the best-fit coefficients. The bivariate regression was performed using the emcee Python package \citep{Foreman_Mackey2013}.\footnote{See details and examples at \url{https://emcee.readthedocs.io/en/stable/}} One advantage of using emcee for regression is that the uncertainties in all of the variables (in this case, 1 dependent and 2 independent variables) can be included in the regression;\footnote{We found the following python tutorial instructive to guide the bivariate fitting in this work: \url{https://dfm.io/posts/fitting-a-plane/}} however, it does not handle censored data. In order to include the small fraction of upper limits in the fitting routine, we again performed resampling of the data 100 times using the methods outlined in the previous sub-section. The best-fit model was chosen to be the model corresponding to the median value of the \heiiew coefficient.\footnote{We also could have chosen the model corresponding to the median coefficient of \Lopt with little difference in the result.} The best-fit equation of the model is the following:

\begin{equation}
\begin{aligned}
\alpha_{\rm ox} = {} & (0.317 \pm 0.040)\mathrm{log(\ion{He}{ii}\ EW)}\\
& - (0.079 \pm 0.017)\mathrm{log}(L_{2500}) + (0.691 \pm 0.555),\\ 
&\sigma_{\epsilon} = (0.086 \pm 0.012).
\end{aligned}
\label{eq:multi}
\end{equation}
Figure \ref{fig:Multivar} depicts \aox as a function of the combination \heiiew and \Lopt using the best-fit model in Equation~\ref{eq:multi}. The solid black line in this figure depicts the $y=x$ line and the grey shaded region depicts the $1\sigma$ confidence interval (computed using the method of \citealt{Kelly2007}). Comparing the coefficients in Equation~\ref{eq:multi}, we find that \heiiew and \Lopt are different from zero by \hbox{$\approx7.9\sigma$} and \hbox{$\approx4.6\sigma$}, respectively, again indicating that both parameters contribute significantly in modeling \aox. As before, this result may suggest that \aox is more dependent on \heiiew than on \Lopt; however, further investigation comparing these parameters is required before any definitive conclusion can be made.

\begin{figure}
\centering
	\includegraphics[width=\columnwidth]{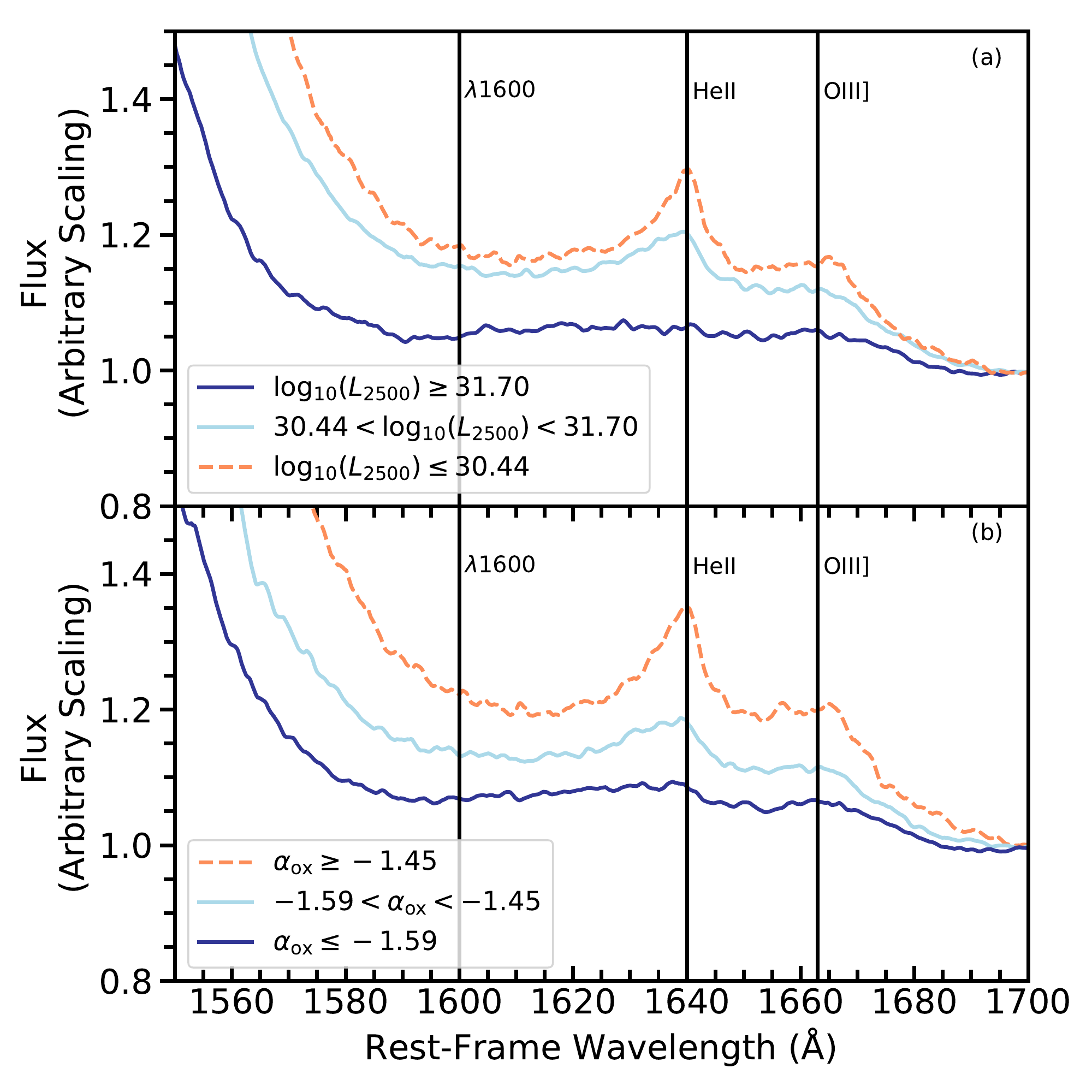}
    \caption{Median-stacked \heii emission-line regions in three bins in ${\rm log}(L_{2500})$ (panel a) and $\alpha_{\rm ox}$ (panel b). As expected from Figure \ref{fig:heii_L2500}, the median \heii emission-line profile of the low-luminosity bin (orange dashed line) in panel (a) is more prominent than both the intermediate-luminosity (light blue line) and high-luminosity (dark blue line) bins. In panel (b), the strength of the median-stacked \heii emission-line profile decreases with decreasing $\alpha_{\rm ox}$ (orange dashed line, light blue line, and dark blue line, respectively) as expected from the relationship depicted in Figure \ref{fig:aox_HeII}. }
    \label{fig:linestack}
\end{figure}

\begin{figure}
\centering
	\includegraphics[width=\columnwidth]{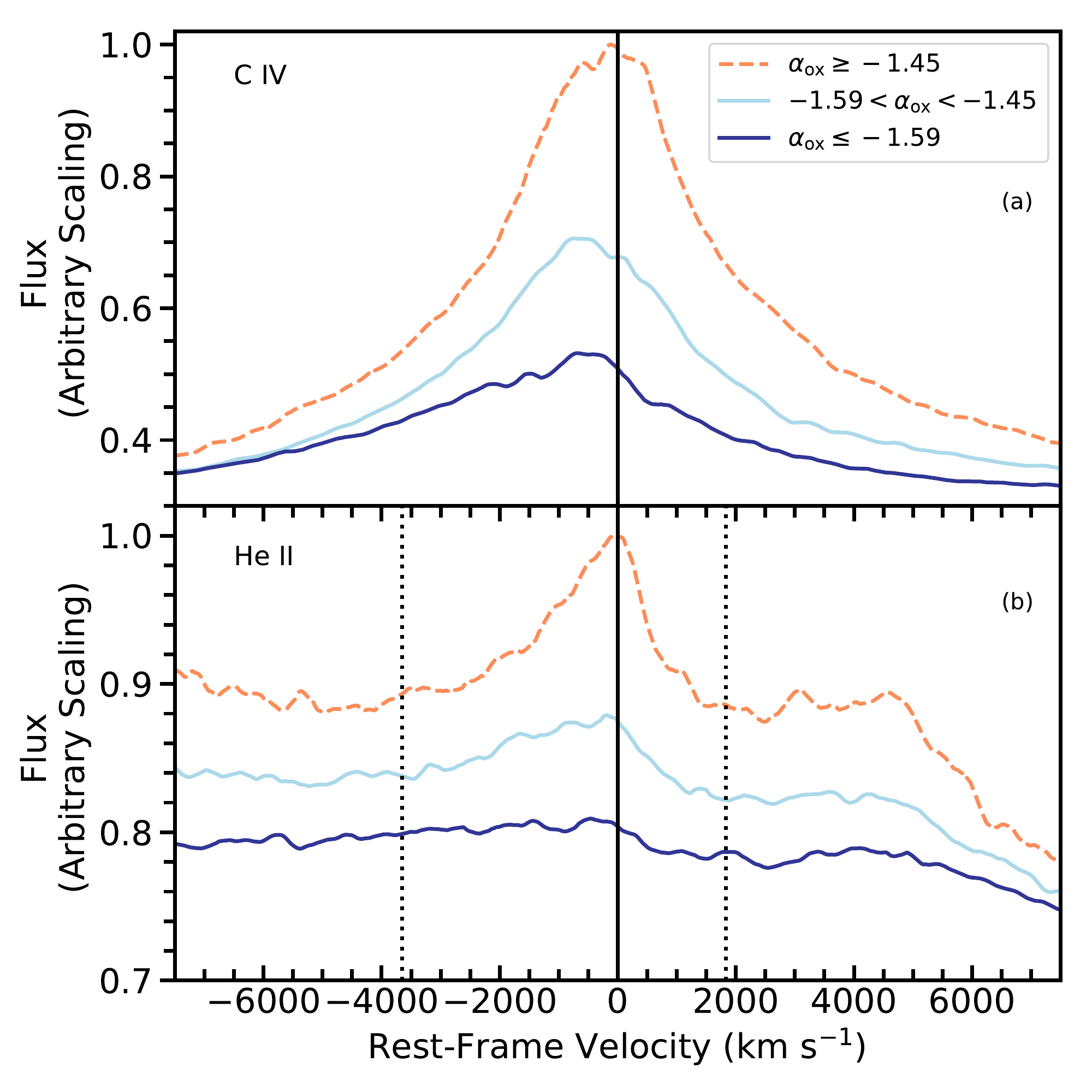}
    \caption{Median-stacked \civ (panel a) and \heii (panel b) emission lines in the three $\alpha_{\rm ox}$ bins from panel (b) of Figure~\ref{fig:linestack}. The solid vertical black lines in both panels depict the vacuum wavelength of the emission-line peaks. The vertical dotted lines in panel (b) depict the region in which the \heii emission significantly contributes to the spectral emission, within which we computed the \heiiew (1620--1650~\AA; see Section~\ref{sec:spec_meas}).  Both panels demonstrate that these emission lines become more blueshifted and increasingly asymmetric about the line center as $\alpha_{\rm ox}$ decreases. The growing asymmetry is seemingly stronger for the \heii emission line within the dotted lines in panel~(b) compared to the evolution seen in the \civ emission line. Panel (b) also demonstrates that the \heii emission line is significantly blended with its surroundings, which motivated our model-independent fitting method. }
    \label{fig:lineprofile}
\end{figure}

A visual comparison of the \heii emission-line profiles of our quasars in bins of \Lopt and \aox, depicted in Figure~\ref{fig:linestack}, also provides useful insights into changes of the \heii emission line with these parameters. To generate panel (a) of Figure \ref{fig:linestack}, we grouped the data first in three bins of \Lopt, where we separated the high-$L$ from the SDSS-RM and PG quasars as one bin, and then split the SDSS-RM and PG by their median \Lopt ($\rm{log}$(\Lopt) = 30.44) to generate the other two lower luminosity bins. Stacking was performed using the cleaned spectra (i.e.\ corrected for Galactic extinction and $3\sigma$ clipped to remove spurious outliers) by projecting the individual spectra in each bin onto a common frame, normalized to the median flux value in the range \hbox{1700--1705 \AA}. The spectra depicted in panel~(a) of Figure \ref{fig:linestack} show the median pixel value of all of the individual spectra within the respective bin. In total, the highest luminosity bin (dark blue) is a stack of 43 spectra, the middle-luminosity bin (light blue) depicts 81 stacked spectra, and the low-luminosity bin (orange dashed) depicts 82 stacked spectra. The spectra in panel~(b) are produced in the same manner; however we binned by \aox, splitting the range of \aox into three, nearly equally-sized bins. The large \aox bin (orange dashed) contains 70 spectra, the middle bin in \aox (light blue) contains 68 spectra, and the small \aox bin (dark blue) also contains 68 spectra. Both panels in Figure \ref{fig:linestack} depict the expected behavior of the \heii emission line, with the low-$L$ bin (and high \aox bin) displaying the largest \heiiew whereas the high-$L$ bin (and small \aox bin) show weak \heii emission. 

We also observe that the \heii emission line is highly asymmetric, and appears to become more asymmetric moving from strong to weak \heii emission, with a blue wing extending as far as $\approx 1590$~\AA\ (corresponding to $\approx 9000$ km s$^{-1}$) at the highest luminosity and steepest \aox bins. Similar behavior is often observed for the \civ emission line as demonstrated in Figure~11 of \citet{Timlin2020} where quasars with weaker \civ emission lines also tend to have more asymmetric and blueshifted line profiles. A visualization of the stacked \civ and \heii emission-line profiles for our sample as a function of \aox is depicted in Figure~\ref{fig:lineprofile}. The \heii emission line rapidly becomes asymmetric with decreasing \aox within the region where it contributes significantly to the spectral emission (i.e. within the dotted lines in panel~(b) of Figure~\ref{fig:lineprofile}). Also, as \aox decreases the line peaks of \civ and \heii appear to be increasingly blueshifted (see Figure~1 in \citealt{Shen2016} for a comparison of the relative line shifts between these lines). Finally, we performed a stacked spectral analysis to assess whether there was a redshift dependence of the \heiiew by comparing the spectral profiles of the PG and RM quasars since they span a similar luminosity range but are at different redshifts. Our data suggest that there is no redshift dependence, where the median \heiiew value for the PG quasars \hbox{(log(\heiiew) $= 0.91 \pm 0.34$)} is statistically consistent with that of the RM quasars \hbox{(log(\heiiew) $= 0.70 \pm 0.20$)}.


\section{Discussion}\label{sec:discussion}

\subsection{Results from earlier related investigations }\label{sec:disc_previous}

Detailed investigations of proxies of the strength of the ionizing EUV continuum have been critically important to modeling and understanding better the physical properties of quasars. For example, the \civ emission line has been shown to exhibit a blueshift with respect to the systemic redshift, which has been used in the literature as observational evidence for a disk-wind model of quasar outflows (e.g.\ \citealt{Richards2011} and references therein). Analyses of the \civew--\civ blueshift parameter space have demonstrated that quasars with large \civew (EW $\geq$ 50 \AA) generally exhibit modest blueshift whereas those with smaller \civew (EW $\leq$ 15 \AA) can exhibit very high-velocity outflows ($v > 4000$~\hbox{km s$^{-1}$}; e.g. \citealt{Wu2011, Luo2015}). In the context of this model, the disparity in the outflow velocity is partly the result of the quasar wind region being over-ionized (in the case of high \civew), and thus the less-energetic UV line driving in this component is inefficient (e.g.\ \citealt{Leighly2004, Richards2011}); however, when the number of ionizing photons in the wind region is small, perhaps due to disk geometry (e.g.\ \citealt{Leighly2004, Luo2015}), the line driving becomes more efficient.

Both \aox and \heiiew have been studied independently to determine their distribution within the \civew--\civ blueshift parameter space and to investigate their relationship with outflow velocity. Investigations of the distribution of \aox in the \civew--\civ blueshift parameter space generally show harder (less negative) \aox values occupy the large \civew, low \civ blueshift region and softer (more negative) \aox values occupy the small \civew, high \civ blueshift space (e.g.\ \citealt{Kruczek2011, Wu2011, Timlin2020}), albeit with significant scatter. Moreover, previous work has demonstrated that \aox is individually correlated with both the \civ emission-line outflow velocity and \civew (e.g.\ \citealt{Vietri2018, Timlin2020}) further indicating that \aox is a good proxy of the balance between the strength of the ionizing SED and the line-driving radiation.

Despite the weakness of the \heii emission line, previous investigations have gathered high-quality spectra with which to measure the \heiiew so that it can be compared with \civ emission-line and absorption-line properties. \citet{Baskin2013, Baskin2015} found that the \heiiew is related to both the profile and blueshift of the high-ionization broad absorption lines (BALs) in quasars, indicating that the \heiiew is a good proxy of the outflow velocity. Recently, \citet{Rankine2020} investigated the behavior of the \heiiew within the \civew--\civ blueshift parameter space (see the left-hand panel of their Figure~12). They found that the average \heiiew smoothly varies throughout the \civew--\civ blueshift parameter space, with large \heiiew values occupying the high \civew, low \civ blueshift region and small \heiiew values occupying the small \civew, large \civ blueshift region of the parameter space. As is the case with \aox, the relationships between the \heiiew and the \civ emission-line properties suggest that the \heiiew is also a useful diagnostic to measure the amount of ionizing radiation and the line-driving efficiency of the less-energetic UV radiation. 

The combination of the results from these previous investigations suggest that a correlation might exist between \aox and \heiiew; however, no direct comparison has been made between these parameters until now.

\subsection{Results of this work}\label{sec:disc_our}

In this work, we found a strong correlation between \aox and \heiiew, both of which can therefore be used as proxies for the strength of the ionizing continuum of the quasar SED. Both of these parameters are also well known to be individually anti-correlated with \Lopt (see the references in Sections~\ref{sec:intro} and \ref{sec:correlations}), and we depict these relationships for our data in Figures~\ref{fig:aox_L2500} and \ref{fig:heii_L2500}. Since both of these parameters exhibit such a strong anti-correlation with \Lopt, we investigated whether or not the \aox--\heiiew correlation is simply a secondary effect resulting from the combination of the \aox--\Lopt and \heiiew--\Lopt relationships. We found that a significant \aox--\heiiew relationship remains after removing the effects of \Lopt (demonstrated in Figures~\ref{fig:Residuals}--\ref{fig:linestack} and by a partial-correlation test). The \heiiew is therefore providing additional independent information into the nature of the ionizing SED in quasars. 

It is not immediately apparent that there should be a strong correlation between \aox and \heiiew since the ionizing radiation responsible for the production of \heii is effectively independent of the harder \xray emission. To demonstrate this mathematically, we adopt a simple power-law to model the ionizing continuum at $\approx 50$~eV (near the ionization energy of \heii) with an energy index of \hbox{$\alpha_{\rm ion} = -1.6$} (e.g.\ Figure~5 in \citealt{Baskin2014}), corresponding to a photon index of $\Gamma_{\rm ion} = -2.6$. Integrating under this power-law indicates that the majority of the photons in the \heii ionizing portion of the SED are largely outside of the soft \xray regime, where \hbox{$\approx$ 90\%} of the photons lie within the energy range of 50 eV to 0.2 keV. The 2 keV \xray emission used to compute \aox, therefore, should have little effect on the production of the \heii ion.\footnote{While we cannot directly measure the monochromatic 2 keV flux density, the median rest-frame energy band utilized in the \heii sample (\hbox{1.68--6.63~keV}) encompasses the rest-frame 2 keV energy and is far above 0.2 keV; see Section~\ref{sec:sample_selection} for the median rest-frame band-pass energies for each individual sample.} Moreover, the ionizing EUV photons are thought to be produced in the inner accretion-disk region (likely with some Comptonization; e.g.\ \citealt{Petrucci2018} and references therein) whereas the harder \xray photons are generally thought to be produced in the corona above the disk. The tight correlation found in this work between \aox and \heiiew therefore suggests that the production mechanism of the harder \xray photons is coupled to that of the EUV photons such that, when the mechanism changes, it affects the strengths of the \xray and EUV portions of the SED similarly. 

The residuals in Figure~\ref{fig:Residuals} demonstrate that the negative correlations of both \aox and \heiiew with \Lopt are significant and non-secondary relationships. The UV continuum is largely generated, in the standard model, by thermal emission from the quasar accretion disk which may not be significantly related to the (non-thermal) mechanism that produces the power-law ionizing EUV continuum. The relationships found in this work, however, suggest that there is a link between the thermal UV emission, the power-law EUV ionizing SED, and the \xray coronal properties in each quasar. In other words, it similarly impacts the strength of the quasar SED in the thermally dominated UV region, the power-law EUV region, and the \xray coronal region. 

The residual \aox--\heiiew relationship (after removing their \Lopt dependences) further demonstrates that these two parameters are related by an additional physical mechanism rather than only the UV luminosity. A physical property which may potentially affect the EUV disk emission, in addition to luminosity, is the gas metallicity. If the commonly observed continuum spectral break below 1000 \AA\ (e.g. \citealt{Lusso2015}) is indeed due to radiation-pressure driven 
mass loss \citep{Laor2014}, then increased metallicity leads to a higher mass-loss rate, which lowers the
inner accretion-disk temperature, and likely lowers the ionizing continuum. The increased disk mass loss, which inevitably passes through the coronal layers, may cool the coronal gas, and thus lower also the \xray emission. 

As outlined in Section~\ref{sec:intro}, the \heiiew is expected to be more direct measure of the strength of the ionizing SED relative to the underlying UV continuum than the \civew. To test this more directly, we compare the strengths of the individual correlations of \civ and \heii with \aox. While multiple studies have investigated the \aox--\civew relationship (e.g.\ \citealt{Green1998, Gibson2008, Kruczek2011, Timlin2020}), we can compare the results above directly with the previous work of \citet{Timlin2020}, in which the \aox--\civew properties of 637 quasars were analyzed, since the analysis methods are nearly identical. We find both a stronger Spearman correlation coefficient and a significantly steeper best-fit slope in the \hbox{\aox--\heiiew} parameter space ($\rho = 0.558 \pm 0.036, \beta = 0.557 \pm 0.039$, respectively; see Table~\ref{tab:Samplestats}) than in the \hbox{\aox--\civew} parameter space ($\rho = 0.470 \pm 0.016, \beta = 0.390 \pm 0.027$, respectively; see Table~2 and Equation~4 in \citealt{Timlin2020}). The differences between the strengths of these relationships reflects the scatter and apparent non-linearity in the relation between the \civew and the \heiiew (see Appendix~\ref{append:C}). The \civew is only a secondary indicator of the ionizing continuum, compared to the \heiiew, as it depends also on the environmental conditions (see Section~\ref{sec:intro}). The relation between the \civew and the \heiiew derived in Appendix~\ref{append:C} can be used to estimate the \heiiew in lower signal-to-noise spectra where only the \civew is measurable.

Our investigation indicates that extreme variability of the ionizing EUV and \xray luminosities are not commonly observed. Our sample was constructed using quasars from multiple different samples that often had large temporal gaps between the observations. Despite the differences in the epoch of observation (generally $> 1$ year), only the low-luminosity quasar Mrk 335 was identified to have extreme \xray variability \citep{Komossa2020}. The extreme \xray variability of Mrk 335 causes this object to be an extreme outlier with respect to the rest of the data, particularly in Figures~\ref{fig:heii_L2500} and \ref{fig:aox_HeII}. Given that only a few quasars exhibit a similarly large scatter as Mrk 335, our investigation suggests that extreme flux variability at the wavelengths of interest in this work is a rare occurrence (consistent with the \citealt{Timlin2020b} direct estimate of the frequency of extreme \xray variability in typical quasars). 

We observed no significant redshift dependence in the quasar SED relationships given the similarity of the PG and SDSS-RM data in Figures~\ref{fig:aox_L2500}--\ref{fig:aox_HeII}. Quasars, therefore, seemingly have a similar mechanism for generating the ionizing SED regardless at which cosmic epoch the quasar resides.

Finally, this work provides important constraints for theoretical models of the quasar SED. In particular, the strong correlations with \Lopt suggest that the spectral shape from 1500~\AA\ to 2~keV is determined primarily by the quasar luminosity. In contrast, the accretion-disk model suggests that the portion of the SED (from 1500 \AA\ to $\approx$50~\AA) which sets the \heiiew should be dependent on the luminosity as well as the mass and spin of the black hole. Additionally, disk-corona models suggest that the 2500~\AA\ to 2~keV SED, which dictates \aox, should be dependent on the coronal optical depth, temperature, and covering factor, in addition to luminosity. If the strength of the SED in these regions is dictated by more than one parameter, one might expect the relationships investigated in this work to exhibit more scatter than observed. Previous work has already found that the \xray and UV emission are tightly correlated, which implies that a physical mechanism regulates the coronal and disk emission (e.g.\ \citealt{Steffen2006, Just2007, Lusso2016, Lusso2017}). The relatively small amount of observed scatter in the overall SED in our work further suggests that there is some unknown regulating mechanism linking the \xray, ionizing EUV, and UV regions of the SED (e.g.\ \citealt{Lawrence2012}). 

One possible interpretation of the EUV--\xray connection is that the \xray emission is generated in a ``hot Comptonization" coronal region and the EUV emission largely originates in a related ``warm Comptonization" coronal region (e.g.\ \citealt{Petrucci2018} and references therein). This scenario, while perhaps aesthetically appealing, currently lacks utility as a predictive model since the relevant hot/warm coronal properties cannot be computed from first principles. Nevertheless, the tight relation of the \xray, EUV, and UV regions of the SED demonstrated in this work provides an important clue that hopefully can be exploited to help reveal the nature of the corona in quasars.


\section{Summary and Future Work}

In this work, we presented the first, large-scale investigation of the joint relationships between \aox, \heiiew, and \Lopt. We gathered 206 quasars that have high-quality spectral coverage of the rest-frame UV and have sensitive \xray coverage with \Chandra or {\emph{XMM-Newton}}; these span wide ranges  of both luminosity and redshift. The main results of the paper are the following:

\begin{itemize}

\item[(i)] We recovered the well-known anti-correlation between \aox and \Lopt (Figure~\ref{fig:aox_L2500}) for our sample, and found consistent best-fit parameters with values presented in previous work. We also investigated the \heii Baldwin effect using the largest sample of measurements from individual quasars to date (as opposed to previous stacked analyses from, e.g.\ \citealt{Dietrich2002}). We found a slightly steeper relation than in previous investigations, confirming that the \heii Baldwin effect is the steepest Baldwin relationship for any emission line. 

\item[(ii)] We found a significant correlation between \aox and \heiiew (Figure~\ref{fig:aox_HeII}) that has a similar correlation strength as the \aox--\Lopt and \heiiew--\Lopt relationships (see Table~\ref{tab:Samplestats}).  The lack of correlations in the residual relationships in Figure~\ref{fig:Residuals} indicates that none of the relationships in Figures~\ref{fig:aox_L2500}--\ref{fig:aox_HeII} is a dominant relationship with the others merely being secondary. The significance of the \heiiew and \Lopt coefficients in the bivariate regression of \aox further confirmed this point (Figure~\ref{fig:Multivar} and Equation~\ref{eq:multi}).

\item[(iii)] A significant correlation was found between \aox and \heiiew even after removing their corresponding luminosity dependences (Section~\ref{sec:prime_second_eff} and Figure~\ref{fig:resid-resid}). This correlation indicates that there is an intrinsic relationship between \aox and \heiiew, the two proxies of the strength/shape of the ionizing continuum investigated in this work. The \heii line is generated largely independently of the \xray continuum, and the two are ultimately produced by radiation originating from different regions of a quasar corona. Therefore, a coupling mechanism must exist that similarly impacts the emission from these regions to form this correlation.

\item[(iv)] The similarity of the SDSS-RM and PG quasars, which have similar luminosity distributions but are located at different redshifts, $z<0.35$ versus $1.6<z<3.5$, in each of the parameter spaces investigated in this work (Figures~\ref{fig:aox_L2500}--\ref{fig:aox_HeII}) indicates that the ionizing continuum has no material redshift dependence. Moreover, the lack of large outliers associated with extremely variable quasars like \hbox{Mrk 335} in our investigation suggests that extreme \xray variability of the quasar luminosity is rare, even across year-long gaps between observations, and thus has little effect on our results. The rarity of extreme \xray variability directly observed in typical quasars further supports this conclusion (e.g.\ \citealt{Timlin2020b}).

\end{itemize}

One way to expand upon the work presented above would be to investigate the distribution of the \xray and \heii properties of other populations of quasars, including radio-loud and BAL quasars. In particular, it would be illuminating to determine if these populations lie on the \aox--\heiiew trend found for the radio-quiet quasars in this work. Their distribution in this space might provide key details regarding the ionizing continuum present in these populations and their relationship to typical quasars. For example, previous work has found that radio-loud quasars tend to have a flatter \xray spectrum (e.g.\ \citealt{Wilkes1987, Reeves1997}), are generally \xray brighter, and have flatter \aox than radio-quiet quasars, which has been used as evidence for a two component model of the \xray emission from radio-loud quasars (e.g.\ \citealt{Worrall1987, Miller2011}). Recent work, however, has demonstrated that this may not be the case generally, and that the \xray emission from most radio-loud quasars might largely be produced in the corona \citep{Zhu2020}. In our work, we found that the \heiiew is an indirect indicator of the strength of the corona, since the \xray emission in radio-quiet quasars largely originates in the corona. Since \heii likely cannot be produced by a radio jet, comparing the \heiiew of a large, representative sample of radio-loud quasars to that of radio-quiet quasars may provide insights regarding the strength of the coronal emission in radio-loud quasars. If the measured \aox and \heiiew values follow the relationship found in our work, then the \xray emission from radio-loud quasars is likely produced mainly in the corona, whereas if there is a large deviation from our relationship, then the jet-linked component likely contributes significantly to the \xray production (a recent investigation can be found in \citealt{Timlin2021}, submitted).

Investigating the relations between the \xray emission, \heiiew, and luminosity of BAL quasars might allow for better constraints on the origin of the BAL phenomenon. Most BAL quasars are found to be \xray absorbed, where studies have found weak soft \xray emission but stronger hard ($> 5$ keV) \xray emission (e.g. \citealt{Gallagher2002, Gallagher2006, Giustini2008, Fan2009}). However, a small fraction of BAL quasars continue to show exceptionally weak hard \xray emission (7--10\%; see \citealt{Liu2018}). One possible explanation for the hard \xray weakness is that such quasars are intrinsically \xray weak \citep{Luo2014}; however, the mechanism causing the hard \xray weakness of these quasars remains unknown. BAL quasars are also known to be preferentially \heii weak, and the fraction of quasars which present BALs increases as the \heiiew decreases (e.g.\ Table~5 of \citealt{Baskin2013}). Given our investigations in this work, weaker \heii emission should be expected for quasars that exhibit weaker \xray emission at a given UV luminosity; however, a quantitative investigation of the \xray and \heii weakness relative to each other might help identify the mechanism causing this weakness. Mapping the location of the known soft \xray absorbed BAL quasars in the \aox--\heiiew parameter space might help constrain better the location of the absorbing material relative to the location of the \xray and EUV emitting regions. For example, stronger \heiiew than expected from the observed \aox might imply that the absorbing material blocks the \xray corona from the observer's line of sight, but does not block the EUV from reaching the \heii broad emission-line region in the quasar. Additionally, investigating the \heiiew properties of the hard \xray weak BAL quasars might also provide insight into the nature of the apparent intrinsic \xray weakness by quantifying the weakness of the EUV continuum. 

Another useful test to perform with these data in future work is a principal component analysis (PCA). PCA can robustly analyze many parameters, and thus along with \heiiew and \Lopt we could investigate the impact of additional physical parameters, such as various emission-line properties. Such an analysis would require additional analysis of these data to obtain the relevant parameters.

\section*{Acknowledgements}

We thank the referee, Andy Lawrence, for his helpful suggestions that improved this manuscript. We thank Jianfeng Wu for helpful discussions. JDT and WNB acknowledge support from NASA ADP grant 80NSSC18K0878, \Chandra \hbox{X-ray} Center grant GO0-21080X, the V.\ M.\ Willaman Endowment, and Penn State ACIS Instrument Team Contract SV4-74018 (issued by the \Chandra \hbox{X-ray} Center, which is operated by the Smithsonian Astrophysical Observatory for and on behalf of NASA under contract NAS8-03060). AL was supported by the Israel Science Foundation (grant no. 1008/18). The \Chandra ACIS Team Guaranteed Time Observations (GTO) utilized were selected by the ACIS Instrument Principal Investigator, Gordon P.\ Garmire, currently of the Huntingdon Institute for \hbox{X-ray} Astronomy, LLC, which is under contract to the Smithsonian Astrophysical Observatory via Contract SV2-82024.

For this research, we have used the Python language along with Astropy\footnote{\url{https://www.astropy.org/}} \citep{astropy2018}, Scipy\footnote{\url{https://www.scipy.org/}} \citep{scipy}, and TOPCAT\footnote{\url{http://www.star.bris.ac.uk/~mbt/topcat/}} \citep{Taylor2005}.

\section*{Data Availability}
The data used in this investigation are available in the article and in its online supplementary material. See Appendices \ref{append:A} and \ref{append:B} for more details.




\bibliographystyle{mnras}
\bibliography{Running_bib} 

\begin{thebibliography}{}
\makeatletter
\relax
\def\mn@urlcharsother{\let\do\@makeother \do\$\do\&\do\#\do\^\do\_\do\%\do\~}
\def\mn@doi{\begingroup\mn@urlcharsother \@ifnextchar [ {\mn@doi@}
  {\mn@doi@[]}}
\def\mn@doi@[#1]#2{\def\@tempa{#1}\ifx\@tempa\@empty \href
  {http://dx.doi.org/#2} {doi:#2}\else \href {http://dx.doi.org/#2} {#1}\fi
  \endgroup}
\def\mn@eprint#1#2{\mn@eprint@#1:#2::\@nil}
\def\mn@eprint@arXiv#1{\href {http://arxiv.org/abs/#1} {{\tt arXiv:#1}}}
\def\mn@eprint@dblp#1{\href {http://dblp.uni-trier.de/rec/bibtex/#1.xml}
  {dblp:#1}}
\def\mn@eprint@#1:#2:#3:#4\@nil{\def\@tempa {#1}\def\@tempb {#2}\def\@tempc
  {#3}\ifx \@tempc \@empty \let \@tempc \@tempb \let \@tempb \@tempa \fi \ifx
  \@tempb \@empty \def\@tempb {arXiv}\fi \@ifundefined
  {mn@eprint@\@tempb}{\@tempb:\@tempc}{\expandafter \expandafter \csname
  mn@eprint@\@tempb\endcsname \expandafter{\@tempc}}}

\bibitem[\protect\citeauthoryear{{Akritas} \& {Siebert}}{{Akritas} \&
  {Siebert}}{1996}]{Akritas1996}
{Akritas} M.~G.,  {Siebert} J.,  1996, \mn@doi [\mnras]
  {10.1093/mnras/278.4.919}, \href
  {https://ui.adsabs.harvard.edu/abs/1996MNRAS.278..919A} {278, 919}

\bibitem[\protect\citeauthoryear{{Antonucci}}{{Antonucci}}{2015}]{Antonucci2015}
{Antonucci} R.,  2015, arXiv e-prints, \href
  {https://ui.adsabs.harvard.edu/abs/2015arXiv150102001A} {p. arXiv:1501.02001}

\bibitem[\protect\citeauthoryear{{Baldwin}}{{Baldwin}}{1977}]{Baldwin1977}
{Baldwin} J.~A.,  1977, \mn@doi [\apj] {10.1086/155294}, \href
  {https://ui.adsabs.harvard.edu/abs/1977ApJ...214..679B} {214, 679}

\bibitem[\protect\citeauthoryear{{Baskin} \& {Laor}}{{Baskin} \&
  {Laor}}{2004}]{Baskin2004}
{Baskin} A.,  {Laor} A.,  2004, \mn@doi [\mnras]
  {10.1111/j.1365-2966.2004.07833.x}, \href
  {https://ui.adsabs.harvard.edu/abs/2004MNRAS.350L..31B} {350, L31}

\bibitem[\protect\citeauthoryear{{Baskin}, {Laor}  \& {Hamann}}{{Baskin}
  et~al.}{2013}]{Baskin2013}
{Baskin} A.,  {Laor} A.,   {Hamann} F.,  2013, \mn@doi [\mnras]
  {10.1093/mnras/stt582}, \href
  {https://ui.adsabs.harvard.edu/abs/2013MNRAS.432.1525B} {432, 1525}

\bibitem[\protect\citeauthoryear{{Baskin}, {Laor}  \& {Stern}}{{Baskin}
  et~al.}{2014}]{Baskin2014}
{Baskin} A.,  {Laor} A.,   {Stern} J.,  2014, \mn@doi [\mnras]
  {10.1093/mnras/stt2230}, \href
  {https://ui.adsabs.harvard.edu/abs/2014MNRAS.438..604B} {438, 604}

\bibitem[\protect\citeauthoryear{{Baskin}, {Laor}  \& {Hamann}}{{Baskin}
  et~al.}{2015}]{Baskin2015}
{Baskin} A.,  {Laor} A.,   {Hamann} F.,  2015, \mn@doi [\mnras]
  {10.1093/mnras/stv406}, \href
  {https://ui.adsabs.harvard.edu/abs/2015MNRAS.449.1593B} {449, 1593}

\bibitem[\protect\citeauthoryear{{Bianchi}, {Guainazzi}, {Matt}, {Fonseca
  Bonilla}  \& {Ponti}}{{Bianchi} et~al.}{2009}]{Bianchi2009}
{Bianchi} S.,  {Guainazzi} M.,  {Matt} G.,  {Fonseca Bonilla} N.,   {Ponti} G.,
   2009, \mn@doi [\aap] {10.1051/0004-6361:200810620}, \href
  {https://ui.adsabs.harvard.edu/abs/2009A&A...495..421B} {495, 421}

\bibitem[\protect\citeauthoryear{{Boroson} \& {Green}}{{Boroson} \&
  {Green}}{1992}]{Boroson1992}
{Boroson} T.~A.,  {Green} R.~F.,  1992, \mn@doi [\apjs] {10.1086/191661}, \href
  {https://ui.adsabs.harvard.edu/abs/1992ApJS...80..109B} {80, 109}

\bibitem[\protect\citeauthoryear{{Brandt}, {Laor}  \& {Wills}}{{Brandt}
  et~al.}{2000}]{Brandt2000}
{Brandt} W.~N.,  {Laor} A.,   {Wills} B.~J.,  2000, \mn@doi [\apj]
  {10.1086/308207}, \href
  {https://ui.adsabs.harvard.edu/abs/2000ApJ...528..637B} {528, 637}

\bibitem[\protect\citeauthoryear{{Chiang} \& {Murray}}{{Chiang} \&
  {Murray}}{1996}]{Chiang1996}
{Chiang} J.,  {Murray} N.,  1996, \mn@doi [\apj] {10.1086/177543}, \href
  {https://ui.adsabs.harvard.edu/abs/1996ApJ...466..704C} {466, 704}

\bibitem[\protect\citeauthoryear{{Davis} \& {Tchekhovskoy}}{{Davis} \&
  {Tchekhovskoy}}{2020}]{Davis2020}
{Davis} S.~W.,  {Tchekhovskoy} A.,  2020, \mn@doi [\araa]
  {10.1146/annurev-astro-081817-051905}, \href
  {https://ui.adsabs.harvard.edu/abs/2020ARA&A..58..407D} {58, 407}

\bibitem[\protect\citeauthoryear{{Dietrich}, {Hamann}, {Shields}, {Constantin},
  {Vestergaard}, {Chaffee}, {Foltz}  \& {Junkkarinen}}{{Dietrich}
  et~al.}{2002}]{Dietrich2002}
{Dietrich} M.,  {Hamann} F.,  {Shields} J.~C.,  {Constantin} A.,  {Vestergaard}
  M.,  {Chaffee} F.,  {Foltz} C.~B.,   {Junkkarinen} V.~T.,  2002, \mn@doi
  [\apj] {10.1086/344410}, \href
  {https://ui.adsabs.harvard.edu/abs/2002ApJ...581..912D} {581, 912}

\bibitem[\protect\citeauthoryear{{Elvis} et~al.,}{{Elvis}
  et~al.}{1994}]{Elvis1994}
{Elvis} M.,  et~al., 1994, \mn@doi [\apjs] {10.1086/192093}, \href
  {https://ui.adsabs.harvard.edu/abs/1994ApJS...95....1E} {95, 1}

\bibitem[\protect\citeauthoryear{{Fan}, {Wang}, {Wang}, {Wang}, {Dong}, {Zhang}
   \& {Cheng}}{{Fan} et~al.}{2009}]{Fan2009}
{Fan} L.~L.,  {Wang} H.~Y.,  {Wang} T.,  {Wang} J.,  {Dong} X.,  {Zhang} K.,
  {Cheng} F.,  2009, \mn@doi [\apj] {10.1088/0004-637X/690/1/1006}, \href
  {https://ui.adsabs.harvard.edu/abs/2009ApJ...690.1006F} {690, 1006}

\bibitem[\protect\citeauthoryear{{Feigelson}, {Nelson}, {Isobe}  \&
  {LaValley}}{{Feigelson} et~al.}{2014}]{Feigelson2014}
{Feigelson} E.~D.,  {Nelson} P.~I.,  {Isobe} T.,   {LaValley} M.,  2014,
  {ASURV: Astronomical SURVival Statistics} (\mn@eprint {ascl} {1406.001})

\bibitem[\protect\citeauthoryear{{Ferland}, {Peterson}, {Horne}, {Welsh}  \&
  {Nahar}}{{Ferland} et~al.}{1992}]{Ferland1992}
{Ferland} G.~J.,  {Peterson} B.~M.,  {Horne} K.,  {Welsh} W.~F.,   {Nahar}
  S.~N.,  1992, \mn@doi [\apj] {10.1086/171063}, \href
  {https://ui.adsabs.harvard.edu/abs/1992ApJ...387...95F} {387, 95}

\bibitem[\protect\citeauthoryear{{Filiz Ak} et~al.,}{{Filiz Ak}
  et~al.}{2012}]{FilizAk2012}
{Filiz Ak} N.,  et~al., 2012, \mn@doi [\apj] {10.1088/0004-637X/757/2/114},
  \href {https://ui.adsabs.harvard.edu/abs/2012ApJ...757..114F} {757, 114}

\bibitem[\protect\citeauthoryear{Foreman-Mackey, Hogg, Lang  \&
  Goodman}{Foreman-Mackey et~al.}{2013}]{Foreman_Mackey2013}
Foreman-Mackey D.,  Hogg D.~W.,  Lang D.,   Goodman J.,  2013, \mn@doi
  [Publications of the Astronomical Society of the Pacific] {10.1086/670067},
  125, 306–312

\bibitem[\protect\citeauthoryear{{Fruscione} et~al.,}{{Fruscione}
  et~al.}{2006}]{Fruscione2006}
{Fruscione} A.,  et~al., 2006, in {Silva} D.~R.,  {Doxsey} R.~E.,  eds,
  Society of Photo-Optical Instrumentation Engineers (SPIE) Conference Series
  Vol. 6270, Society of Photo-Optical Instrumentation Engineers (SPIE)
  Conference Series. p. 62701V, \mn@doi{10.1117/12.671760}

\bibitem[\protect\citeauthoryear{{Gallagher}, {Brandt}, {Chartas}  \&
  {Garmire}}{{Gallagher} et~al.}{2002}]{Gallagher2002}
{Gallagher} S.~C.,  {Brandt} W.~N.,  {Chartas} G.,   {Garmire} G.~P.,  2002,
  \mn@doi [\apj] {10.1086/338485}, \href
  {https://ui.adsabs.harvard.edu/abs/2002ApJ...567...37G} {567, 37}

\bibitem[\protect\citeauthoryear{{Gallagher}, {Brandt}, {Chartas}, {Priddey},
  {Garmire}  \& {Sambruna}}{{Gallagher} et~al.}{2006}]{Gallagher2006}
{Gallagher} S.~C.,  {Brandt} W.~N.,  {Chartas} G.,  {Priddey} R.,  {Garmire}
  G.~P.,   {Sambruna} R.~M.,  2006, \mn@doi [\apj] {10.1086/503762}, \href
  {https://ui.adsabs.harvard.edu/abs/2006ApJ...644..709G} {644, 709}

\bibitem[\protect\citeauthoryear{{Gibson}, {Brandt}  \& {Schneider}}{{Gibson}
  et~al.}{2008}]{Gibson2008}
{Gibson} R.~R.,  {Brandt} W.~N.,   {Schneider} D.~P.,  2008, \mn@doi [\apj]
  {10.1086/590403}, \href
  {https://ui.adsabs.harvard.edu/abs/2008ApJ...685..773G} {685, 773}

\bibitem[\protect\citeauthoryear{{Giustini}, {Cappi}  \& {Vignali}}{{Giustini}
  et~al.}{2008}]{Giustini2008}
{Giustini} M.,  {Cappi} M.,   {Vignali} C.,  2008, \mn@doi [\aap]
  {10.1051/0004-6361:200810363}, \href
  {https://ui.adsabs.harvard.edu/abs/2008A&A...491..425G} {491, 425}

\bibitem[\protect\citeauthoryear{{Goad} \& {Korista}}{{Goad} \&
  {Korista}}{2014}]{Goad2014}
{Goad} M.~R.,  {Korista} K.~T.,  2014, \mn@doi [\mnras]
  {10.1093/mnras/stu1456}, \href
  {https://ui.adsabs.harvard.edu/abs/2014MNRAS.444...43G} {444, 43}

\bibitem[\protect\citeauthoryear{{Green}}{{Green}}{1996}]{Green1996}
{Green} P.~J.,  1996, \mn@doi [\apj] {10.1086/177584}, \href
  {https://ui.adsabs.harvard.edu/abs/1996ApJ...467...61G} {467, 61}

\bibitem[\protect\citeauthoryear{{Green}}{{Green}}{1998}]{Green1998}
{Green} P.~J.,  1998, \mn@doi [\apj] {10.1086/305537}, \href
  {https://ui.adsabs.harvard.edu/abs/1998ApJ...498..170G} {498, 170}

\bibitem[\protect\citeauthoryear{{Green}, {Schmidt}  \& {Liebert}}{{Green}
  et~al.}{1986}]{Green1986}
{Green} R.~F.,  {Schmidt} M.,   {Liebert} J.,  1986, \mn@doi [\apjs]
  {10.1086/191115}, \href
  {https://ui.adsabs.harvard.edu/abs/1986ApJS...61..305G} {61, 305}

\bibitem[\protect\citeauthoryear{{Guo}, {Shen}  \& {Wang}}{{Guo}
  et~al.}{2018}]{Guo2018}
{Guo} H.,  {Shen} Y.,   {Wang} S.,  2018, {PyQSOFit: Python code to fit the
  spectrum of quasars}, Astrophysics Source Code Library (\mn@eprint {ascl}
  {1809.008})

\bibitem[\protect\citeauthoryear{{Inoue}, {Terashima}  \& {Ho}}{{Inoue}
  et~al.}{2007}]{Inoue2007}
{Inoue} H.,  {Terashima} Y.,   {Ho} L.~C.,  2007, \mn@doi [\apj]
  {10.1086/517995}, \href
  {https://ui.adsabs.harvard.edu/abs/2007ApJ...662..860I} {662, 860}

\bibitem[\protect\citeauthoryear{Jones, Oliphant, Peterson  et~al.}{Jones
  et~al.}{2001}]{scipy}
Jones E.,  Oliphant T.,  Peterson P.,   et~al., 2001, {SciPy}: Open source
  scientific tools for {Python}, \url {http://www.scipy.org/}

\bibitem[\protect\citeauthoryear{{Just}, {Brandt}, {Shemmer}, {Steffen},
  {Schneider}, {Chartas}  \& {Garmire}}{{Just} et~al.}{2007}]{Just2007}
{Just} D.~W.,  {Brandt} W.~N.,  {Shemmer} O.,  {Steffen} A.~T.,  {Schneider}
  D.~P.,  {Chartas} G.,   {Garmire} G.~P.,  2007, \mn@doi [\apj]
  {10.1086/519990}, \href
  {https://ui.adsabs.harvard.edu/abs/2007ApJ...665.1004J} {665, 1004}

\bibitem[\protect\citeauthoryear{{Kalberla}, {Burton}, {Hartmann}, {Arnal},
  {Bajaja}, {Morras}  \& {P{\"o}ppel}}{{Kalberla} et~al.}{2005}]{Kalberla2005}
{Kalberla} P.~M.~W.,  {Burton} W.~B.,  {Hartmann} D.,  {Arnal} E.~M.,  {Bajaja}
  E.,  {Morras} R.,   {P{\"o}ppel} W.~G.~L.,  2005, \mn@doi [\aap]
  {10.1051/0004-6361:20041864}, \href
  {https://ui.adsabs.harvard.edu/abs/2005A&A...440..775K} {440, 775}

\bibitem[\protect\citeauthoryear{{Kaspi}, {Maoz}, {Netzer}, {Peterson},
  {Vestergaard}  \& {Jannuzi}}{{Kaspi} et~al.}{2005}]{Kaspi2005}
{Kaspi} S.,  {Maoz} D.,  {Netzer} H.,  {Peterson} B.~M.,  {Vestergaard} M.,
  {Jannuzi} B.~T.,  2005, \mn@doi [\apj] {10.1086/431275}, \href
  {https://ui.adsabs.harvard.edu/abs/2005ApJ...629...61K} {629, 61}

\bibitem[\protect\citeauthoryear{{Kellermann}, {Sramek}, {Schmidt}, {Green}  \&
  {Shaffer}}{{Kellermann} et~al.}{1994}]{Kellermann1994}
{Kellermann} K.~I.,  {Sramek} R.~A.,  {Schmidt} M.,  {Green} R.~F.,   {Shaffer}
  D.~B.,  1994, \mn@doi [\aj] {10.1086/117145}, \href
  {https://ui.adsabs.harvard.edu/abs/1994AJ....108.1163K} {108, 1163}

\bibitem[\protect\citeauthoryear{{Kelly}}{{Kelly}}{2007}]{Kelly2007}
{Kelly} B.~C.,  2007, \mn@doi [\apj] {10.1086/519947}, \href
  {https://ui.adsabs.harvard.edu/abs/2007ApJ...665.1489K} {665, 1489}

\bibitem[\protect\citeauthoryear{Kendall}{Kendall}{1938}]{Kendall1938}
Kendall M.~G.,  1938, Biometrika, 30, 81

\bibitem[\protect\citeauthoryear{Kendall}{Kendall}{1970}]{Kendall1970}
Kendall M.,  1970, Rank Correlation Methods.
Theory and applications of rank order-statistics, Griffin, \url
  {https://books.google.com/books?id=Mm2jjgEACAAJ}

\bibitem[\protect\citeauthoryear{{Komossa} et~al.,}{{Komossa}
  et~al.}{2020}]{Komossa2020}
{Komossa} S.,  et~al., 2020, \mn@doi [\aap] {10.1051/0004-6361/202039098},
  \href {https://ui.adsabs.harvard.edu/abs/2020A&A...643L...7K} {643, L7}

\bibitem[\protect\citeauthoryear{{Korista}, {Baldwin}  \& {Ferland}}{{Korista}
  et~al.}{1998}]{Korista1998}
{Korista} K.,  {Baldwin} J.,   {Ferland} G.,  1998, \mn@doi [\apj]
  {10.1086/306321}, \href
  {https://ui.adsabs.harvard.edu/abs/1998ApJ...507...24K} {507, 24}

\bibitem[\protect\citeauthoryear{{Kraft}, {Burrows}  \& {Nousek}}{{Kraft}
  et~al.}{1991}]{Kraft1991}
{Kraft} R.~P.,  {Burrows} D.~N.,   {Nousek} J.~A.,  1991, \mn@doi [\apj]
  {10.1086/170124}, \href
  {https://ui.adsabs.harvard.edu/abs/1991ApJ...374..344K} {374, 344}

\bibitem[\protect\citeauthoryear{{Krawczyk}, {Richards}, {Mehta}, {Vogeley},
  {Gallagher}, {Leighly}, {Ross}  \& {Schneider}}{{Krawczyk}
  et~al.}{2013}]{Krawczyk2013}
{Krawczyk} C.~M.,  {Richards} G.~T.,  {Mehta} S.~S.,  {Vogeley} M.~S.,
  {Gallagher} S.~C.,  {Leighly} K.~M.,  {Ross} N.~P.,   {Schneider} D.~P.,
  2013, \mn@doi [\apjs] {10.1088/0067-0049/206/1/4}, \href
  {https://ui.adsabs.harvard.edu/abs/2013ApJS..206....4K} {206, 4}

\bibitem[\protect\citeauthoryear{{Kruczek} et~al.,}{{Kruczek}
  et~al.}{2011}]{Kruczek2011}
{Kruczek} N.~E.,  et~al., 2011, \mn@doi [\aj] {10.1088/0004-6256/142/4/130},
  \href {https://ui.adsabs.harvard.edu/abs/2011AJ....142..130K} {142, 130}

\bibitem[\protect\citeauthoryear{{Laor} \& {Brandt}}{{Laor} \&
  {Brandt}}{2002}]{Laor2002}
{Laor} A.,  {Brandt} W.~N.,  2002, \mn@doi [\apj] {10.1086/339476}, \href
  {https://ui.adsabs.harvard.edu/abs/2002ApJ...569..641L} {569, 641}

\bibitem[\protect\citeauthoryear{{Laor} \& {Davis}}{{Laor} \&
  {Davis}}{2014}]{Laor2014}
{Laor} A.,  {Davis} S.~W.,  2014, \mn@doi [\mnras] {10.1093/mnras/stt2408},
  \href {https://ui.adsabs.harvard.edu/abs/2014MNRAS.438.3024L} {438, 3024}

\bibitem[\protect\citeauthoryear{{Laor}, {Bahcall}, {Jannuzi}, {Schneider}  \&
  {Green}}{{Laor} et~al.}{1995}]{Laor1995}
{Laor} A.,  {Bahcall} J.~N.,  {Jannuzi} B.~T.,  {Schneider} D.~P.,   {Green}
  R.~F.,  1995, \mn@doi [\apjs] {10.1086/192177}, \href
  {https://ui.adsabs.harvard.edu/abs/1995ApJS...99....1L} {99, 1}

\bibitem[\protect\citeauthoryear{{Lawrence}}{{Lawrence}}{2012}]{Lawrence2012}
{Lawrence} A.,  2012, \mn@doi [\mnras] {10.1111/j.1365-2966.2012.20889.x},
  \href {https://ui.adsabs.harvard.edu/abs/2012MNRAS.423..451L} {423, 451}

\bibitem[\protect\citeauthoryear{{Lawrence}}{{Lawrence}}{2018}]{Lawrence2018}
{Lawrence} A.,  2018, \mn@doi [Nature Astronomy] {10.1038/s41550-017-0372-1},
  \href {https://ui.adsabs.harvard.edu/abs/2018NatAs...2..102L} {2, 102}

\bibitem[\protect\citeauthoryear{{Leighly}}{{Leighly}}{2004}]{Leighly2004}
{Leighly} K.~M.,  2004, \mn@doi [\apj] {10.1086/422089}, \href
  {https://ui.adsabs.harvard.edu/abs/2004ApJ...611..125L} {611, 125}

\bibitem[\protect\citeauthoryear{{Liu}, {Luo}, {Brandt}, {Gallagher}  \&
  {Garmire}}{{Liu} et~al.}{2018}]{Liu2018}
{Liu} H.,  {Luo} B.,  {Brandt} W.~N.,  {Gallagher} S.~C.,   {Garmire} G.~P.,
  2018, \mn@doi [\apj] {10.3847/1538-4357/aabe8d}, \href
  {https://ui.adsabs.harvard.edu/abs/2018ApJ...859..113L} {859, 113}

\bibitem[\protect\citeauthoryear{{Liu} et~al.,}{{Liu} et~al.}{2020}]{Liu2020}
{Liu} T.,  et~al., 2020, \mn@doi [\apjs] {10.3847/1538-4365/abb5b0}, \href
  {https://ui.adsabs.harvard.edu/abs/2020ApJS..250...32L} {250, 32}

\bibitem[\protect\citeauthoryear{Liu, Luo, Brandt, Brotherton, Gallagher, Ni,
  Shemmer  \& au2}{Liu et~al.}{2021}]{Liu2021}
Liu H.,  Luo B.,  Brandt W.~N.,  Brotherton M.~S.,  Gallagher S.~C.,  Ni Q.,
  Shemmer O.,   au2 J. D. T.~I.,  2021, On the Observational Difference Between
  the Accretion Disk-Corona Connections among Super- and Sub-Eddington
  Accreting Active Galactic Nuclei (\mn@eprint {arXiv} {2102.02832})

\bibitem[\protect\citeauthoryear{{Luo} et~al.,}{{Luo} et~al.}{2014}]{Luo2014}
{Luo} B.,  et~al., 2014, \mn@doi [\apj] {10.1088/0004-637X/794/1/70}, \href
  {https://ui.adsabs.harvard.edu/abs/2014ApJ...794...70L} {794, 70}

\bibitem[\protect\citeauthoryear{{Luo} et~al.,}{{Luo} et~al.}{2015}]{Luo2015}
{Luo} B.,  et~al., 2015, \mn@doi [\apj] {10.1088/0004-637X/805/2/122}, \href
  {https://ui.adsabs.harvard.edu/abs/2015ApJ...805..122L} {805, 122}

\bibitem[\protect\citeauthoryear{{Lusso} \& {Risaliti}}{{Lusso} \&
  {Risaliti}}{2016}]{Lusso2016}
{Lusso} E.,  {Risaliti} G.,  2016, \mn@doi [\apj]
  {10.3847/0004-637X/819/2/154}, \href
  {https://ui.adsabs.harvard.edu/abs/2016ApJ...819..154L} {819, 154}

\bibitem[\protect\citeauthoryear{{Lusso} \& {Risaliti}}{{Lusso} \&
  {Risaliti}}{2017}]{Lusso2017}
{Lusso} E.,  {Risaliti} G.,  2017, \mn@doi [\aap]
  {10.1051/0004-6361/201630079}, \href
  {https://ui.adsabs.harvard.edu/abs/2017A&A...602A..79L} {602, A79}

\bibitem[\protect\citeauthoryear{{Lusso}, {Worseck}, {Hennawi}, {Prochaska},
  {Vignali}, {Stern}  \& {O'Meara}}{{Lusso} et~al.}{2015}]{Lusso2015}
{Lusso} E.,  {Worseck} G.,  {Hennawi} J.~F.,  {Prochaska} J.~X.,  {Vignali} C.,
   {Stern} J.,   {O'Meara} J.~M.,  2015, \mn@doi [\mnras]
  {10.1093/mnras/stv516}, \href
  {https://ui.adsabs.harvard.edu/abs/2015MNRAS.449.4204L} {449, 4204}

\bibitem[\protect\citeauthoryear{Lyons}{Lyons}{1991}]{Lyons1991}
Lyons L.,  1991, A Practical Guide to Data Analysis for Physical Science
  Students.
Cambridge University Press

\bibitem[\protect\citeauthoryear{{MacLeod} et~al.,}{{MacLeod}
  et~al.}{2010}]{MacLeod2010}
{MacLeod} C.~L.,  et~al., 2010, \mn@doi [\apj] {10.1088/0004-637X/721/2/1014},
  \href {https://ui.adsabs.harvard.edu/abs/2010ApJ...721.1014M} {721, 1014}

\bibitem[\protect\citeauthoryear{{Miller}, {Brandt}, {Schneider}, {Gibson},
  {Steffen}  \& {Wu}}{{Miller} et~al.}{2011}]{Miller2011}
{Miller} B.~P.,  {Brandt} W.~N.,  {Schneider} D.~P.,  {Gibson} R.~R.,
  {Steffen} A.~T.,   {Wu} J.,  2011, \mn@doi [\apj]
  {10.1088/0004-637X/726/1/20}, \href
  {https://ui.adsabs.harvard.edu/abs/2011ApJ...726...20M} {726, 20}

\bibitem[\protect\citeauthoryear{{Netzer}}{{Netzer}}{2013}]{Netzer2013}
{Netzer} H.,  2013, {The Physics and Evolution of Active Galactic Nuclei}

\bibitem[\protect\citeauthoryear{{Neugebauer}, {Green}, {Matthews}, {Schmidt},
  {Soifer}  \& {Bennett}}{{Neugebauer} et~al.}{1987}]{Neugebauer1987}
{Neugebauer} G.,  {Green} R.~F.,  {Matthews} K.,  {Schmidt} M.,  {Soifer}
  B.~T.,   {Bennett} J.,  1987, \mn@doi [\apjs] {10.1086/191175}, \href
  {https://ui.adsabs.harvard.edu/abs/1987ApJS...63..615N} {63, 615}

\bibitem[\protect\citeauthoryear{{Ni} et~al.,}{{Ni} et~al.}{2018}]{Ni2018}
{Ni} Q.,  et~al., 2018, \mn@doi [\mnras] {10.1093/mnras/sty1989}, \href
  {https://ui.adsabs.harvard.edu/abs/2018MNRAS.480.5184N} {480, 5184}

\bibitem[\protect\citeauthoryear{{P{\^a}ris} et~al.,}{{P{\^a}ris}
  et~al.}{2018}]{Paris2018}
{P{\^a}ris} I.,  et~al., 2018, \mn@doi [\aap] {10.1051/0004-6361/201732445},
  \href {https://ui.adsabs.harvard.edu/abs/2018A&A...613A..51P} {613, A51}

\bibitem[\protect\citeauthoryear{{Petrucci}, {Ursini}, {De Rosa}, {Bianchi},
  {Cappi}, {Matt}, {Dadina}  \& {Malzac}}{{Petrucci}
  et~al.}{2018}]{Petrucci2018}
{Petrucci} P.~O.,  {Ursini} F.,  {De Rosa} A.,  {Bianchi} S.,  {Cappi} M.,
  {Matt} G.,  {Dadina} M.,   {Malzac} J.,  2018, \mn@doi [\aap]
  {10.1051/0004-6361/201731580}, \href
  {https://ui.adsabs.harvard.edu/abs/2018A&A...611A..59P} {611, A59}

\bibitem[\protect\citeauthoryear{{Piconcelli}, {Jimenez-Bail{\'o}n},
  {Guainazzi}, {Schartel}, {Rodr{\'\i}guez-Pascual}  \&
  {Santos-Lle{\'o}}}{{Piconcelli} et~al.}{2005}]{Piconcelli2005}
{Piconcelli} E.,  {Jimenez-Bail{\'o}n} E.,  {Guainazzi} M.,  {Schartel} N.,
  {Rodr{\'\i}guez-Pascual} P.~M.,   {Santos-Lle{\'o}} M.,  2005, \mn@doi [\aap]
  {10.1051/0004-6361:20041621}, \href
  {https://ui.adsabs.harvard.edu/abs/2005A&A...432...15P} {432, 15}

\bibitem[\protect\citeauthoryear{Press, Teukolsky, Vetterling  \&
  Flannery}{Press et~al.}{1993}]{Press1992}
Press W.~H.,  Teukolsky S.~A.,  Vetterling W.~T.,   Flannery B.~P.,  1993,
  Numerical Recipes in FORTRAN; The Art of Scientific Computing, 2nd edn.
Cambridge University Press, 40 W. 20 St. New York, NY, USA

\bibitem[\protect\citeauthoryear{{Price-Whelan} et~al.,}{{Price-Whelan}
  et~al.}{2018}]{astropy2018}
{Price-Whelan} A.~M.,  et~al., 2018, \mn@doi [\aj] {10.3847/1538-3881/aabc4f},
  \href {https://ui.adsabs.harvard.edu/#abs/2018AJ....156..123T} {156, 123}

\bibitem[\protect\citeauthoryear{{Pu}, {Luo}, {Brandt}, {Timlin}, {Liu}, {Ni}
  \& {Wu}}{{Pu} et~al.}{2020}]{Pu2020}
{Pu} X.,  {Luo} B.,  {Brandt} W.~N.,  {Timlin} J.~D.,  {Liu} H.,  {Ni} Q.,
  {Wu} J.,  2020, \mn@doi [\apj] {10.3847/1538-4357/abacc5}, \href
  {https://ui.adsabs.harvard.edu/abs/2020ApJ...900..141P} {900, 141}

\bibitem[\protect\citeauthoryear{{Rankine}, {Hewett}, {Banerji}  \&
  {Richards}}{{Rankine} et~al.}{2020}]{Rankine2020}
{Rankine} A.~L.,  {Hewett} P.~C.,  {Banerji} M.,   {Richards} G.~T.,  2020,
  \mn@doi [\mnras] {10.1093/mnras/staa130}, \href
  {https://ui.adsabs.harvard.edu/abs/2020MNRAS.492.4553R} {492, 4553}

\bibitem[\protect\citeauthoryear{{Reeves}, {Turner}, {Ohashi}  \&
  {Kii}}{{Reeves} et~al.}{1997}]{Reeves1997}
{Reeves} J.~N.,  {Turner} M.~J.~L.,  {Ohashi} T.,   {Kii} T.,  1997, \mn@doi
  [\mnras] {10.1093/mnras/292.3.468}, \href
  {https://ui.adsabs.harvard.edu/abs/1997MNRAS.292..468R} {292, 468}

\bibitem[\protect\citeauthoryear{{Richards} et~al.,}{{Richards}
  et~al.}{2006}]{Richards2006}
{Richards} G.~T.,  et~al., 2006, \mn@doi [\aj] {10.1086/503559}, \href
  {https://ui.adsabs.harvard.edu/abs/2006AJ....131.2766R} {131, 2766}

\bibitem[\protect\citeauthoryear{{Richards} et~al.,}{{Richards}
  et~al.}{2011}]{Richards2011}
{Richards} G.~T.,  et~al., 2011, \mn@doi [\aj] {10.1088/0004-6256/141/5/167},
  \href {https://ui.adsabs.harvard.edu/abs/2011AJ....141..167R} {141, 167}

\bibitem[\protect\citeauthoryear{{Rivera}, {Richards}, {Hewett}  \&
  {Rankine}}{{Rivera} et~al.}{2020}]{Rivera2020}
{Rivera} A.~B.,  {Richards} G.~T.,  {Hewett} P.~C.,   {Rankine} A.~L.,  2020,
  \mn@doi [\apj] {10.3847/1538-4357/aba62c}, \href
  {https://ui.adsabs.harvard.edu/abs/2020ApJ...899...96R} {899, 96}

\bibitem[\protect\citeauthoryear{{Schmidt} \& {Green}}{{Schmidt} \&
  {Green}}{1983}]{Schmidt1983}
{Schmidt} M.,  {Green} R.~F.,  1983, \mn@doi [\apj] {10.1086/161048}, \href
  {https://ui.adsabs.harvard.edu/abs/1983ApJ...269..352S} {269, 352}

\bibitem[\protect\citeauthoryear{{Schneider} et~al.,}{{Schneider}
  et~al.}{2005}]{Schneider2005}
{Schneider} D.~P.,  et~al., 2005, \mn@doi [\aj] {10.1086/431156}, \href
  {https://ui.adsabs.harvard.edu/abs/2005AJ....130..367S} {130, 367}

\bibitem[\protect\citeauthoryear{{Schneider} et~al.,}{{Schneider}
  et~al.}{2010}]{Schneider2010}
{Schneider} D.~P.,  et~al., 2010, \mn@doi [\aj] {10.1088/0004-6256/139/6/2360},
  \href {https://ui.adsabs.harvard.edu/abs/2010AJ....139.2360S} {139, 2360}

\bibitem[\protect\citeauthoryear{{Scott}, {Kriss}, {Brotherton}, {Green},
  {Hutchings}, {Shull}  \& {Zheng}}{{Scott} et~al.}{2004}]{Scott2004}
{Scott} J.~E.,  {Kriss} G.~A.,  {Brotherton} M.,  {Green} R.~F.,  {Hutchings}
  J.,  {Shull} J.~M.,   {Zheng} W.,  2004, \mn@doi [\apj] {10.1086/422336},
  \href {https://ui.adsabs.harvard.edu/abs/2004ApJ...615..135S} {615, 135}

\bibitem[\protect\citeauthoryear{{Shemmer} \& {Lieber}}{{Shemmer} \&
  {Lieber}}{2015}]{Shemmer2015}
{Shemmer} O.,  {Lieber} S.,  2015, \mn@doi [\apj]
  {10.1088/0004-637X/805/2/124}, \href
  {https://ui.adsabs.harvard.edu/abs/2015ApJ...805..124S} {805, 124}

\bibitem[\protect\citeauthoryear{{Shemmer}, {Brandt}, {Netzer}, {Maiolino}  \&
  {Kaspi}}{{Shemmer} et~al.}{2008}]{Shemmer2008}
{Shemmer} O.,  {Brandt} W.~N.,  {Netzer} H.,  {Maiolino} R.,   {Kaspi} S.,
  2008, \mn@doi [\apj] {10.1086/588776}, \href
  {https://ui.adsabs.harvard.edu/abs/2008ApJ...682...81S} {682, 81}

\bibitem[\protect\citeauthoryear{{Shen} et~al.,}{{Shen}
  et~al.}{2011}]{Shen2011}
{Shen} Y.,  et~al., 2011, \mn@doi [\apjs] {10.1088/0067-0049/194/2/45}, \href
  {https://ui.adsabs.harvard.edu/abs/2011ApJS..194...45S} {194, 45}

\bibitem[\protect\citeauthoryear{{Shen} et~al.,}{{Shen}
  et~al.}{2015}]{Shen2015}
{Shen} Y.,  et~al., 2015, \mn@doi [\apjs] {10.1088/0067-0049/216/1/4}, \href
  {https://ui.adsabs.harvard.edu/abs/2015ApJS..216....4S} {216, 4}

\bibitem[\protect\citeauthoryear{{Shen} et~al.,}{{Shen}
  et~al.}{2016}]{Shen2016}
{Shen} Y.,  et~al., 2016, \mn@doi [\apj] {10.3847/0004-637X/831/1/7}, \href
  {https://ui.adsabs.harvard.edu/abs/2016ApJ...831....7S} {831, 7}

\bibitem[\protect\citeauthoryear{{Shen} et~al.,}{{Shen}
  et~al.}{2019}]{Shen2019}
{Shen} Y.,  et~al., 2019, \mn@doi [\apjs] {10.3847/1538-4365/ab074f}, \href
  {https://ui.adsabs.harvard.edu/abs/2019ApJS..241...34S} {241, 34}

\bibitem[\protect\citeauthoryear{{Smee} et~al.,}{{Smee}
  et~al.}{2013}]{Smee2013}
{Smee} S.~A.,  et~al., 2013, \mn@doi [\aj] {10.1088/0004-6256/146/2/32}, \href
  {https://ui.adsabs.harvard.edu/abs/2013AJ....146...32S} {146, 32}

\bibitem[\protect\citeauthoryear{{Steffen}, {Strateva}, {Brandt}, {Alexander},
  {Koekemoer}, {Lehmer}, {Schneider}  \& {Vignali}}{{Steffen}
  et~al.}{2006}]{Steffen2006}
{Steffen} A.~T.,  {Strateva} I.,  {Brandt} W.~N.,  {Alexander} D.~M.,
  {Koekemoer} A.~M.,  {Lehmer} B.~D.,  {Schneider} D.~P.,   {Vignali} C.,
  2006, \mn@doi [\aj] {10.1086/503627}, \href
  {https://ui.adsabs.harvard.edu/abs/2006AJ....131.2826S} {131, 2826}

\bibitem[\protect\citeauthoryear{{Stevans}, {Shull}, {Danforth}  \&
  {Tilton}}{{Stevans} et~al.}{2014}]{Stevans2014}
{Stevans} M.~L.,  {Shull} J.~M.,  {Danforth} C.~W.,   {Tilton} E.~M.,  2014,
  \mn@doi [\apj] {10.1088/0004-637X/794/1/75}, \href
  {https://ui.adsabs.harvard.edu/abs/2014ApJ...794...75S} {794, 75}

\bibitem[\protect\citeauthoryear{{Taylor}}{{Taylor}}{2005}]{Taylor2005}
{Taylor} M.~B.,  2005, in {Shopbell} P.,  {Britton} M.,   {Ebert} R.,  eds,
  Astronomical Society of the Pacific Conference Series Vol. 347, Astronomical
  Data Analysis Software and Systems XIV. p.~29

\bibitem[\protect\citeauthoryear{{Timlin}, {Brandt}, {Ni}, {Luo}, {Pu},
  {Schneider}, {Vivek}  \& {Yi}}{{Timlin} et~al.}{2020a}]{Timlin2020}
{Timlin} J.~D.,  {Brandt} W.~N.,  {Ni} Q.,  {Luo} B.,  {Pu} X.,  {Schneider}
  D.~P.,  {Vivek} M.,   {Yi} W.,  2020a, \mn@doi [\mnras]
  {10.1093/mnras/stz3433}, \href
  {https://ui.adsabs.harvard.edu/abs/2020MNRAS.492..719T} {492, 719}

\bibitem[\protect\citeauthoryear{{Timlin}, {Brandt}, {Zhu}, {Liu}, {Luo}  \&
  {Ni}}{{Timlin} et~al.}{2020b}]{Timlin2020b}
{Timlin} John~D. I.,  {Brandt} W.~N.,  {Zhu} S.,  {Liu} H.,  {Luo} B.,   {Ni}
  Q.,  2020b, \mn@doi [\mnras] {10.1093/mnras/staa2661}, \href
  {https://ui.adsabs.harvard.edu/abs/2020MNRAS.498.4033T} {498, 4033}

\bibitem[\protect\citeauthoryear{{Timlin}, {Zhu}, {Brandt}  \& {Laor}}{{Timlin}
  et~al.}{2021}]{Timlin2021}
{Timlin} J.~D.,  {Zhu} S.~F.,  {Brandt} W.~N.,   {Laor} A.,  2021, RNAAS,
  Submitted

\bibitem[\protect\citeauthoryear{{Vestergaard} \& {Wilkes}}{{Vestergaard} \&
  {Wilkes}}{2001}]{Vestergaard2001}
{Vestergaard} M.,  {Wilkes} B.~J.,  2001, \mn@doi [\apjs] {10.1086/320357},
  \href {https://ui.adsabs.harvard.edu/abs/2001ApJS..134....1V} {134, 1}

\bibitem[\protect\citeauthoryear{{Vietri} et~al.,}{{Vietri}
  et~al.}{2018}]{Vietri2018}
{Vietri} G.,  et~al., 2018, \mn@doi [\aap] {10.1051/0004-6361/201732335}, \href
  {https://ui.adsabs.harvard.edu/abs/2018A&A...617A..81V} {617, A81}

\bibitem[\protect\citeauthoryear{{Waters}, {Kashi}, {Proga}, {Eracleous},
  {Barth}  \& {Greene}}{{Waters} et~al.}{2016}]{Waters2016}
{Waters} T.,  {Kashi} A.,  {Proga} D.,  {Eracleous} M.,  {Barth} A.~J.,
  {Greene} J.,  2016, \mn@doi [\apj] {10.3847/0004-637X/827/1/53}, \href
  {https://ui.adsabs.harvard.edu/abs/2016ApJ...827...53W} {827, 53}

\bibitem[\protect\citeauthoryear{{Webb} et~al.,}{{Webb}
  et~al.}{2020}]{Webb2020}
{Webb} N.~A.,  et~al., 2020, \mn@doi [\aap] {10.1051/0004-6361/201937353},
  \href {https://ui.adsabs.harvard.edu/abs/2020A&A...641A.136W} {641, A136}

\bibitem[\protect\citeauthoryear{{Wilkes} \& {Elvis}}{{Wilkes} \&
  {Elvis}}{1987}]{Wilkes1987}
{Wilkes} B.~J.,  {Elvis} M.,  1987, \mn@doi [\apj] {10.1086/165822}, \href
  {https://ui.adsabs.harvard.edu/abs/1987ApJ...323..243W} {323, 243}

\bibitem[\protect\citeauthoryear{{Worrall}, {Giommi}, {Tananbaum}  \&
  {Zamorani}}{{Worrall} et~al.}{1987}]{Worrall1987}
{Worrall} D.~M.,  {Giommi} P.,  {Tananbaum} H.,   {Zamorani} G.,  1987, \mn@doi
  [\apj] {10.1086/164999}, \href
  {https://ui.adsabs.harvard.edu/abs/1987ApJ...313..596W} {313, 596}

\bibitem[\protect\citeauthoryear{{Wu} et~al.,}{{Wu} et~al.}{2011}]{Wu2011}
{Wu} J.,  et~al., 2011, \mn@doi [\apj] {10.1088/0004-637X/736/1/28}, \href
  {https://ui.adsabs.harvard.edu/abs/2011ApJ...736...28W} {736, 28}

\bibitem[\protect\citeauthoryear{{York} et~al.,}{{York}
  et~al.}{2000}]{York2000}
{York} D.~G.,  et~al., 2000, \mn@doi [\aj] {10.1086/301513}, \href
  {https://ui.adsabs.harvard.edu/abs/2000AJ....120.1579Y} {120, 1579}

\bibitem[\protect\citeauthoryear{{Zheng} \& {Malkan}}{{Zheng} \&
  {Malkan}}{1993}]{Zheng1993}
{Zheng} W.,  {Malkan} M.~A.,  1993, \mn@doi [\apj] {10.1086/173182}, \href
  {https://ui.adsabs.harvard.edu/abs/1993ApJ...415..517Z} {415, 517}

\bibitem[\protect\citeauthoryear{{Zhu}, {Brandt}, {Luo}, {Wu}, {Xue}  \&
  {Yang}}{{Zhu} et~al.}{2020}]{Zhu2020}
{Zhu} S.~F.,  {Brandt} W.~N.,  {Luo} B.,  {Wu} J.,  {Xue} Y.~Q.,   {Yang} G.,
  2020, \mn@doi [\mnras] {10.1093/mnras/staa1411}, \href
  {https://ui.adsabs.harvard.edu/abs/2020MNRAS.496..245Z} {496, 245}

\makeatother
\end{thebibliography}

\appendix
\section{The \ion{He}{ii} Sample}\label{append:A}
We report the measurements of \aox, \heiiew, and \Lopt in Table~\ref{tab:heii_sample} for the 206 quasars that comprise the \ion{He}{ii} sample used to generate the figures in this work. Columns (1) and (2) of Table~\ref{tab:heii_sample} report the sample from which the quasar was selected (\highL, SDSS-RM, PG; see Section~\ref{sec:sample_selection}) and an identification (ID) for each quasar, where the \highL quasars are designated by the SDSS name, the SDSS-RM quasars are identified by their RMID (see \citealt{Shen2019}), and the PG quasars are designated by their PG name. Columns~(3)--(5) report the position (RA and Dec in J2000 degrees) and redshift of the quasars in the sample. The measured log$_{10}$(\heiiew) (in units of \AA), corresponding $1\sigma$ uncertainty (measurement uncertainty only; see Section~\ref{sec:spec_meas} on how to incorporate the variability uncertainty), and a flag indicating whether the \heii emission line is detected ({\tt{True}}=detected; {\tt{False}}=not detected) are reported in columns \hbox{(6)--(8)}. In cases where the \heii emission is not detected, we report the upper limit on the \heiiew (see Section~\ref{sec:spec_meas} for details) and the median uncertainty of the sample from which the quasar was drawn (e.g.\ \highL, SDSS-RM, or PG). Column (9) reports the logarithm of \Lopt (in units of erg s$^{-1}$ Hz$^{-1}$). The measurement uncertainty of \Lopt is not a significant contributor to the overall uncertainty when incorporating the variability and thus we exclude it from this Table (see Section~\ref{sec:sample_selection} on how to compute the variability uncertainty). Columns (10) and (11) report the logarithm of $L_{2 \rm keV}$ and its measurement uncertainty. We report the \aox value, corresponding uncertainty from \Lopt and $L_{2 \rm keV}$, and a flag indicating whether the quasar was \xray-detected in columns~\hbox{(12)--(14)}. In cases where the quasar is not \xray-detected, the upper limit on \aox is reported in column~(12) and the median uncertainty is reported as before in column~(13). Finally, we report the logarithm of the \civew and the uncertainty on that value in columns~\hbox{(15)--(16)}. We describe how we obtained these values for \civ in more detail in Appendix~\ref{append:C}. The full table is available online in machine-readable format.

\begin{table}\label{tab:heii_sample}
\begin{adjustbox}{addcode={\begin{minipage}{\width}}{\caption{
The \heii sample of 206 quasars used for the analysis in this work. Appendix \ref{append:A} presents the full table schema and describes each column in this table in more detail. The full table is available online in machine-readable format. 
$^{*}$ We report the log of the value. 
$^{**}$ We report the uncertainty of the log value. 
}\end{minipage}},rotate=90,center,width=0.45\textwidth}

\begin{tabular}{|r|r|r|r|c|r|c|c|c|c|c|c|c|c|c|c|}
\hline
  \multicolumn{1}{|c|}{Sample} &
  \multicolumn{1}{|c|}{ID} &
  \multicolumn{1}{|c|}{RA} &
  \multicolumn{1}{|c|}{DEC} &
  \multicolumn{1}{|c|}{$z$} &
  \multicolumn{1}{|c|}{\heiiew$^*$} &
  \multicolumn{1}{|c|}{err\_HeII$^{**}$} &
  \multicolumn{1}{|c|}{Hdet} &
  \multicolumn{1}{|c|}{\Lopt$^*$} &
  \multicolumn{1}{|c|}{$L_{\rm 2 keV}^*$} &
  \multicolumn{1}{|c|}{err\_L2keV$^{**}$} &
  \multicolumn{1}{|c|}{\aox}  &
  \multicolumn{1}{|c|}{err\_\aox} &
  \multicolumn{1}{|c|}{Xdet} &
  \multicolumn{1}{|c|}{\civew$^*$} &
  \multicolumn{1}{|c|}{err\_CIV$^{**}$}\\
  
  \multicolumn{1}{|c|}{} &
  \multicolumn{1}{|c|}{} &
  \multicolumn{1}{|c|}{(J2000 deg)} &
  \multicolumn{1}{|c|}{(J2000 deg)} &
  \multicolumn{1}{|c|}{} &
  \multicolumn{1}{|c|}{\AA} &
  \multicolumn{1}{|c|}{} &
  \multicolumn{1}{|c|}{} &
  \multicolumn{1}{|c|}{erg s$^{-1}$ Hz$^{-1}$} &
  \multicolumn{1}{|c|}{erg s$^{-1}$ Hz$^{-1}$} &
  \multicolumn{1}{|c|}{} &
  \multicolumn{1}{|c|}{} &
  \multicolumn{1}{|c|}{} &
  \multicolumn{1}{|c|}{} &
  \multicolumn{1}{|c|}{\AA} &
  \multicolumn{1}{|c|}{} \\
  
  \multicolumn{1}{|c|}{(1)} &
  \multicolumn{1}{|c|}{(2)} &
  \multicolumn{1}{|c|}{(3)} &
  \multicolumn{1}{|c|}{(4)} &
  \multicolumn{1}{|c|}{(5)} &
  \multicolumn{1}{|c|}{(6)} &
  \multicolumn{1}{|c|}{(7)} &
  \multicolumn{1}{|c|}{(8)} &
  \multicolumn{1}{|c|}{(9)} &
  \multicolumn{1}{|c|}{(10)} &
  \multicolumn{1}{|c|}{(11)} &
  \multicolumn{1}{|c|}{(12)} &
  \multicolumn{1}{|c|}{(13)} &
  \multicolumn{1}{|c|}{(14)} &
  \multicolumn{1}{|c|}{(15)} &
  \multicolumn{1}{|c|}{(16)} \\
  
\hline
  High-$L$ & \begin{tabular}{@{}c@{}}012156.03 \\ +144823.9\end{tabular} & 20.483 & 14.806 & 2.864 & 0.459 & 0.114 & 1 & 32.090 & 27.629 & 0.075 & $-$1.710 & 0.028 & 1 & 1.482 & 0.009 \\
  High-$L$ & \begin{tabular}{@{}c@{}}012412.46 \\ $-$010049.8\end{tabular}  & 21.051 & $-$1.013 & 2.903 & $-$0.207 & 0.116 & 0 & 32.064 & 26.873 & 0.301 & $-$1.992 & 0.045 & 0 & 1.480 & 0.005 \\
  SDSS-RM & 8 & 213.478 & 53.006 & 1.612 & 0.915 & 0.177 & 1 & 30.196 & 26.743 & 0.085 & $-$1.325 & 0.032 & 1 & 1.770 & 0.030 \\
  SDSS-RM & 11 & 213.892 & 52.962 & 2.052 & 0.611 & 0.053 & 1 & 30.506 & 26.689 & 0.129 & $-$1.465 & 0.049 & 1 & 1.962 & 0.009\\
  PG & 1116+215 & 169.786 & 21.321 & 0.176 & 0.271 & 0.301 & 1 & 30.656 & 26.728 & 0.010 & $-$1.507 & 0.003 & 1 & 1.852 & 0.052 \\
  PG & 1202+281 & 181.175 & 27.903 & 0.165 & 1.317 & 0.151 & 1 & 29.778 & 26.423 & 0.008 & $-$1.287 & 0.003 & 1 & 1.976 & 0.064 \\
\hline
\end{tabular}

\end{adjustbox}
\end{table}



\section{Cycle 13 \Chandra snapshot observation data reduction}\label{append:B}
In this appendix, we provide a description of the results from the {\emph{Chandra}} Cycle~13 snapshot observations of the luminous quasars from SDSS DR7 used, in part, to construct our \highL sample in Section~\ref{sec:sample_selection}. First, we present the target selection followed by an overview of the \xray image-reduction and source-detection methods used to analyze the {\emph{Chandra}} observations (see Section~3.1 of \citealt{Timlin2020} for more details). Additionally, we present a catalog of the \xray properties of the 26 quasars used in the \highL sample in Table~\ref{tab:schema}, and provide descriptions of the columns in the schema below.

The quasars targeted in this program were selected to be used in conjunction with the quasar targets in \citet{Just2007} to measure the \xray properties of the most-luminous SDSS quasars. The target quasars were required to have an $i$-band absolute magnitude brighter than $M_i = -29.02$ and were not previously observed in either \citet{Just2007} or serendipitously with {\emph{Chandra, XMM-Newton, Swift, ROSAT}}, or {\emph{Einstein}}. These restrictions yielded 66 quasars, of which 65 were ultimately observed and, of these, 26 satisfied our criteria to be used in this work. Exposure times were set at $\approx 1.5$~ks.

The raw data from each {\emph{Chandra}} observation were downloaded from ChaSeR\footnote{\url{https://cda.harvard.edu/chaser/}}, and each observation was reprocessed and filtered for background flaring using the built-in routines in the {\emph{Chandra}} Interactive Analysis of Observations (CIAO; \citealt{Fruscione2006}) software. Images of the soft (0.5--2 keV), hard (2--7 keV), and full (0.5--7 keV) energy bands were produced for each observation, and sources were detected using {\tt{wavdetect}} for all three images. Circular regions of radius 2\arcsec\ centered at the \xray source position nearest to, but no greater than 1\arcsec\ from, the optically determined quasar position from the DR7Q catalog \citep{Schneider2010} were used to extract counts from the images. Background counts were measured in a 50\arcsec\ circular, source-free region near the quasar position. The binomial no-source probability was computed for each energy band using the source and background counts, as well as the ratio of the aperture areas (see Equation~1 in \citealt{Timlin2020}). The quasars were considered to be \xray detected in a band if the binomial probability was $P_B<0.01$, indicating that there was $<1$\% chance of the source being random background events. This threshold is appropriate for targets of pre-specified position, and it was confirmed through visual inspection of the images. Band ratios, effective power-law photon-index values, and soft-band flux values were computed in the same manner as detailed in Section~3.1 of \citet{Timlin2020}. If the quasar was detected in both the soft and hard bands the computed effective power-law photon-index values (which have an average of $\langle\Gamma\rangle = 1.88$) were used to compute the \xray flux, otherwise $\Gamma = 1.90$ was adopted. The observed flux density at rest-frame 2~keV was computed for each quasar using the soft-band fluxes and was used along with the observed flux density at rest-frame 2500~\AA\ (from \citealt{Shen2011}) to compute \aox. Below, we present the description of the columns in the \xray catalog released as part of this work.

\begin{table*}
\centering
\caption{Cycle 13 \Chandra Snapshot Observations}
\label{tab:schema}

\begin{tabular}{|l|r|r|r|r|r|r|r|r|}
\hline
  \multicolumn{1}{|c|}{Name} &
  \multicolumn{1}{c|}{RA} &
  \multicolumn{1}{c|}{DEC} &
  \multicolumn{1}{c|}{$z$} &
  \multicolumn{1}{c|}{ObsID} &
  \multicolumn{1}{c|}{Full\_cts} &
  \multicolumn{1}{c|}{$f_{2 \rm keV}$} &
  \multicolumn{1}{c|}{$f_{2500}$} &
  \multicolumn{1}{c|}{$\alpha_{\rm ox}$}  \\
  
  \multicolumn{1}{|c|}{} &
  \multicolumn{1}{c|}{(J2000 deg)} &
  \multicolumn{1}{c|}{(J2000 deg)} &
  \multicolumn{1}{c|}{} &
  \multicolumn{1}{c|}{} &
  \multicolumn{1}{c|}{} &
  \multicolumn{1}{c|}{erg cm$^{-2}$  s$^{-1}$ Hz$^{-1}$} &
  \multicolumn{1}{c|}{erg cm$^{-2}$  s$^{-1}$ Hz$^{-1}$} &
  \multicolumn{1}{c|}{} \\
  
  \multicolumn{1}{|c|}{(1)} &
  \multicolumn{1}{c|}{(2)} &
  \multicolumn{1}{c|}{(3)} &
  \multicolumn{1}{c|}{(4)} &
  \multicolumn{1}{c|}{(8)} &
  \multicolumn{1}{c|}{(28)} &
  \multicolumn{1}{c|}{(52)} &
  \multicolumn{1}{c|}{(55)} &
  \multicolumn{1}{c|}{(57)} \\
  
\hline
030341.04$-$002321.9	& 45.9210 & $-$0.3890 & 3.234 &	13349 & 9.5756 & 4.55E$-$32 &	4.55E$-$27 &	$-$1.91 \\
084846.10+611234.6 & 132.1920 & 61.2099 & 4.376 & 13353 & 43.7574 & 4.67E$-$31 & 8.07E$-$27	& $-$1.59 \\
114358.52+052444.9 & 175.9940 & 5.4120 & 2.566 & 13368 & 19.1730 & 1.71E$-$31 & 6.29E$-$27 & $-$1.75 \\
160441.47+164538.3	 & 241.1730 & 16.7609 & 2.942 & 13311 & 12.7510 & 4.33E$-$32 & 9.59E$-$27 & $-$2.05 \\
\hline
\end{tabular}

\begin{flushleft}
\footnotesize{{\it Notes:} Select columns from the \xray catalog of the Cycle 13 quasars used in this work. The \xray information of the 26 Cycle~13 targets is presented in this catalog. Details regarding the \xray reduction methods and schema of this table can be found in Appendix~\ref{append:B}. The full table is available online in machine-readable format.}
\end{flushleft}

\end{table*}

\begin{itemize}

\item[--] Column (1): SDSS Name
\item[--] Column (2): J2000 Right Ascension (J2000 degrees)
\item[--] Column (3): J2000 Declination (J2000 degrees)
\item[--] Column (4): Redshift (see \citealt{Shen2011})
\item[--] Column (5): Galactic column density (cm$^{-2}$; \citealt{Kalberla2005})
\item[--] Column (6): Observed $i$-band magnitude 
\item[--] Column (7): Absolute $i$-band magnitude (corrected to $z=2$; \citealt{Richards2006})
\item[--] Column (8): \Chandra observation ID
\item[--] Column (9): \Chandra off-axis angle (arcmin)
\item[--] Column (10): Soft-band effective exposure time (seconds)
\item[--] Column (11): Hard-band effective exposure time (seconds)
\item[--] Column (12): Full-band effective exposure time (seconds)
\item[--] Column (13)--(15): Binomial probability of detection (soft-band, hard-band, full-band)
\item[--] Column (16)--(17): Raw source and background counts (soft-band)
\item[--] Column (18)--(20): Soft-band net counts (or 90\% confidence upper limits); lower and upper $1\sigma$ uncertainty 
\item[--] Column (21)--(22): Raw source and background counts (hard-band)
\item[--] Column (23)--(25): Hard-band net counts (or 90\% confidence upper limits); lower and upper $1\sigma$ uncertainty
\item[--] Column (26)--(27): Raw source and background counts (full-band)
\item[--] Column (28)--(30): Full-band net counts (or 90\% confidence upper limits); lower and upper $1\sigma$ uncertainty
\item[--] Column (31)--(32): Mean exposure-map pixel value of the source and background regions (cm$^2$ s; soft band)
\item[--] Column (33)--(34): Mean exposure-map pixel value of the source and background regions (cm$^2$ s; hard band)
\item[--] Column (35)--(36): Mean exposure-map pixel value of the source and background regions (cm$^2$ s; full band)
\item[--] Column (37)--(39): Hardness ratio and lower and upper limits
\item[--] Column (40)--(42): Power-law photon index (dual-band detections) or limit (single-band detection) and lower/upper limits
\item[--] Column (43)--(45): Soft-band count rate (cts s$^{-1}$); lower and upper $1\sigma$ uncertainty
\item[--] Column (46)--(48): Hard-band count rate (cts s$^{-1}$); lower and upper $1\sigma$ uncertainty
\item[--] Column (49)--(51): Full-band count rate (cts s$^{-1}$); lower and upper $1\sigma$ uncertainty
\item[--] Column (52)--(54): Observed flux density at rest-frame 2 keV \mbox{(erg cm$^{-2}$  s$^{-1}$ Hz$^{-1}$)}; lower and upper limit
\item[--] Column (55): Observed flux density at rest-frame 2500 \angstrom \mbox{(erg cm$^{-2}$  s$^{-1}$ Hz$^{-1}$)}
\item[--] Column (56): Rest-frame 2500 \angstrom monochromatic luminosity (erg s$^{-1}$ Hz$^{-1}$)
\item[--] Column (57)--(59): Observed \aox; lower and upper limit

\end{itemize}


\section{The \heiiew--\civew relation}\label{append:C}

We have investigated the relationship between the \heiiew and the \civew for the quasars in our \heii sample. The measurement of the \civew for the \highL and SDSS-RM quasars was performed consistently using the PyQSOFit fitting software \citep{Guo2018}, which performs Gaussian model fitting to the emission-line profile. For this analysis, we fit the \civ emission line in the same way as described in Section~3.2 of \citet{Timlin2020}. After correcting for Galactic absorption, each spectrum was normalized by a local continuum, where we adopted the wavelength regions from Section~\ref{sec:spec_meas} as the local continuum. We then fit a three-Gaussian model to the \civ emission line within the range $\lambda=$ 1500--1600 \AA. The \civew was determined by integrating this model. The uncertainty in the \civew was determined using a similar Monte Carlo approach to that described in Section~\ref{sec:spec_meas}. Since PyQSOFit is tailored to fit SDSS spectra, we instead measured the \civew for the PG quasars following the method in Section~\ref{sec:spec_meas}, changing the integration window to $\lambda=$ 1500--1600 \AA\ for \civ. We tested this method for consistency with PyQSOFit using the \highL and SDSS-RM quasars and found acceptable agreement between methods, where the average percentage difference between the \civew computed using the two different methods is $\approx 5$\%. 

\begin{figure}
\centering
	\includegraphics[width=\columnwidth]{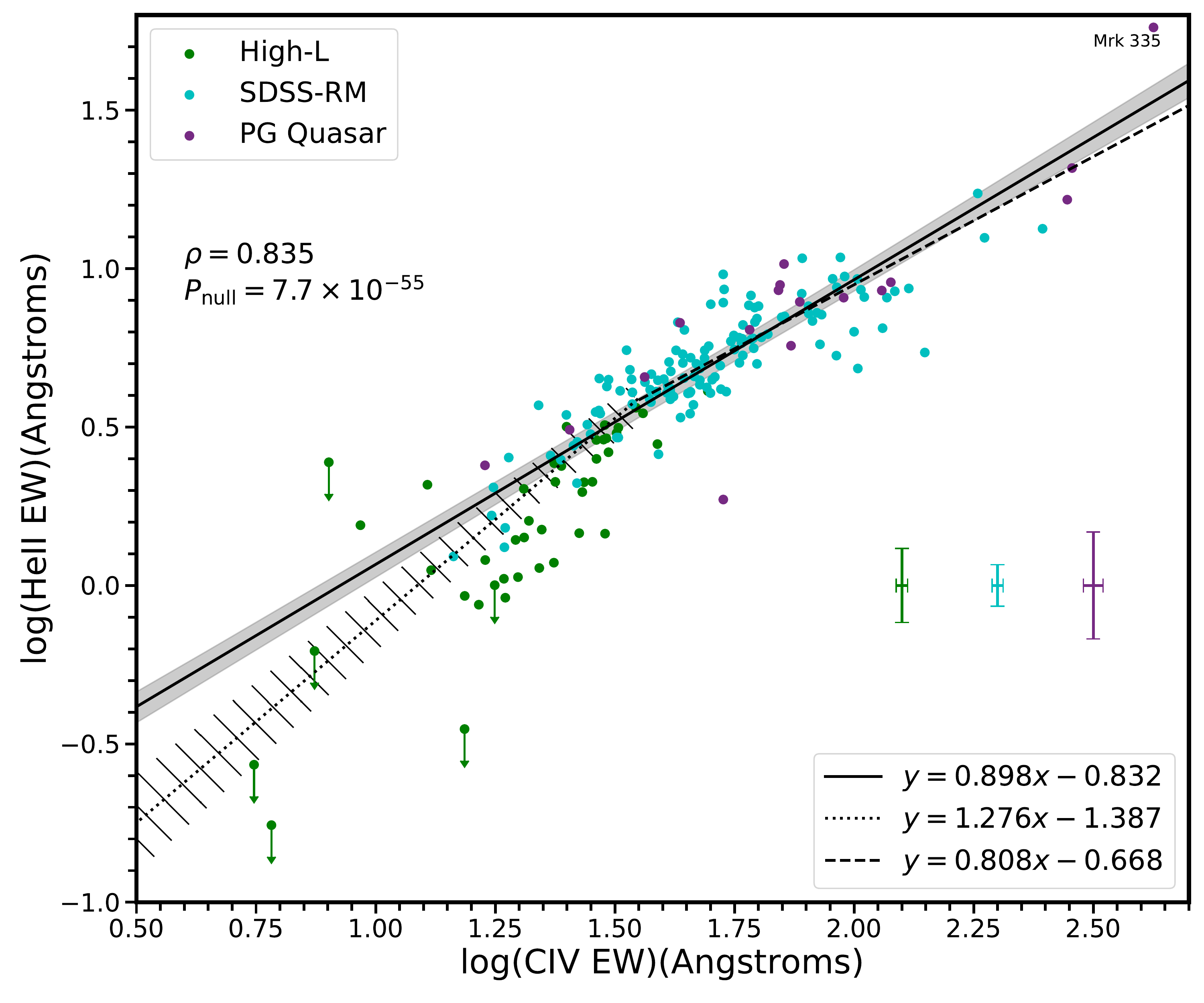}
    \caption{The \heiiew as a function of \civew for the quasars in our \heii sample. The colors and symbols of the points are the same as in Figure~\ref{fig:aox_L2500}. We find a strong correlation between these line EWs (Spearman statistics provided in the top left). We fit a power-law model (solid black line and grey region) to these data using the method from \citet{Kelly2007} and present the functional form in the lower right-hand corner. The \civew is a good indicator of the \heiiew at large values \hbox{(log(\civew) $> 1.55$)}; however, the relationship seemingly diverges from the linear fit at low \civew. We fit a power-law model to the data with \hbox{log(\civew) $< 1.55$} (black dotted line, hatched confidence interval) and to the data with \hbox{log(\civew) $> 1.55$} (dashed solid line, no confidence interval is shown since it overlaps well with the fit to the full data). The fit to the low \civew data clearly diverges from the other two fits. This divergence might be the consequence of the quasar environment that affects \civ, as discussed in Section~\ref{sec:intro}.  }
    \label{fig:CIV_HeII}
\end{figure}

We depict the \heiiew as a function of \civew in Figure~\ref{fig:CIV_HeII}. A Spearman test indicates that there is a strong correlation between the two parameters ($\rho = 0.835$), and thus we fit a power-law relationship using the method from \citet{Kelly2007}. We find the best-fit relationship to be
\begin{equation}\label{eqn:C}
\begin{split}
\civew = (0.898 \pm 0.0354) \rm{log}(\heiiew) \\- (0.832 \pm 0.060); \sigma_i = 0.075 \pm 0.032,
\end{split}
\end{equation}
where $\sigma_i$ is the intrinsic dispersion in the relationship. Given that the observed scatter is $\sigma_o = 0.138$, the uncertainty in the measurements, mainly in \heiiew, accounts for a large fraction of the observed scatter. The tightness of the relationship is clearly demonstrated in Figure~\ref{fig:CIV_HeII} at large values of \civew where the power-law well represents the data. As the \civew decreases, however, more scatter is apparent in the data around the model. In particular, at \hbox{log$_{10}$(\civew) $< 1.55$ \AA}, the data seemingly diverge from the power-law model. To quantify this non-linearity, we fit two additional power-law models to the data, one to the data points with \hbox{log$_{10}$(\civew) $< 1.55$ \AA} (dotted line in Figure~\ref{fig:CIV_HeII}, region represents the 1$\sigma$ confidence interval) and the other with \hbox{log$_{10}$(\civew) $> 1.55$ \AA} (dashed black line). The equations of these fits are provided in Figure~\ref{fig:CIV_HeII}, and the intrinsic scatter of the data around the fits to the low and high \civew data are $\sigma_i = 0.093 \pm 0.056$ and $\sigma_i = 0.074 \pm 0.035$, respectively. The low \civew power-law clearly exhibits a divergence from the rest of the data, and the fit suggests that the intrinsic scatter is mildly larger (albeit within $1\sigma$ of the other fits).

This apparent non-linear trend at low \civew might be a consequence of the environmental effects discussed in Section~\ref{sec:intro} and \ref{sec:disc_our} that affect \civew but not \heiiew. The small intrinsic scatter in this relationship further indicates that the measured \civew is generally a good proxy for the strength of the ionizing continuum, particularly at large values of \civew, and thus is a useful diagnostic of the quasar EUV ionizing SED in low signal-to-noise spectra where \heii is difficult to measure. Aside from the mild non-linearity, this tight relationship indicates that the effects of the quasar environment upon \civew are not typically strong, and theoretical models of quasar environments must be able to reproduce this behavior. Finally, given that most quasars do not have high-quality \xray coverage, the two power-law model enables us to predict the strength of the ionizing continuum from \civew without having to compute \aox. To assess how well this model can constrain the \heiiew, we separated the data into small and large \civew bins by splitting at \hbox{log(\ion{C}{iv} EW) $=1.55$}. We then input the data from the small and large \civew bins into the respective \heiiew--\civew relationship presented at the bottom right-hand side of Figure~\ref{fig:CIV_HeII} to predict \heiiew. The residuals of the prediction to the measured \heiiew were computed, and we found a $1\sigma$ spread in the residual to be $\approx$ 0.132. Since the residuals are calculated using logarithmic values, the difference can be expressed as the logarithm of the ratio of these values. Taking the inverse logarithm of the spread of the residuals indicates that the two power-law model can constrain the \heiiew within a factor of $\approx 1.36$.

\bsp	
\label{lastpage}
\end{document}